\pdfoutput=1

\documentclass{article}

\usepackage{arxiv}

\usepackage{amsmath}

\usepackage[utf8]{inputenc} 
\usepackage[T1]{fontenc}    
\usepackage{hyperref}       
\usepackage{url}            
\usepackage{booktabs}       
\usepackage{amsfonts}       
\usepackage{nicefrac}       
\usepackage{microtype}      
\usepackage{cleveref}       
\usepackage{lipsum}         
\usepackage{graphicx}
\usepackage{natbib}
\usepackage{doi}
\usepackage{mfirstuc}

\usepackage[english]{babel} 
\usepackage{amssymb}
\usepackage{mathdots}
\usepackage[classicReIm]{kpfonts}

\usepackage{mwe,cuted}
\usepackage{float}
\usepackage{subfig}
\usepackage{makecell} 
\usepackage{multirow} 
\newcommand{\reffig}[1]{Figure \ref{#1}}

\title{High-Accuracy Absolute-Position-Aided Code Phase Tracking Based on RTK/INS Deep Integration in Challenging Static Scenarios}

\author{ \href{https://orcid.org/0000-0002-8112-4376}{\includegraphics[scale=0.06]{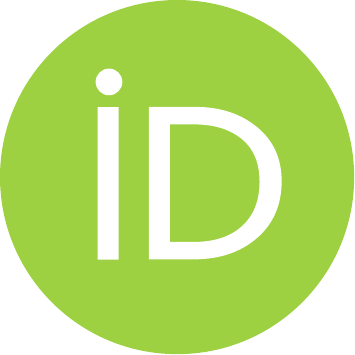}\hspace{1mm}Yiran~Luo}\\
	Department of Geomatics Engineering\\
	University of Calgary\\
	Calgary T2N1N4, Canada \\
	\texttt{yiran.luo@ucalgary.ca} \\
	\And
	\href{https://orcid.org/0000-0002-0352-741X}{\includegraphics[scale=0.06]{orcid.pdf}\hspace{1mm}Li-Ta~Hsu} \\
	Department of Aeronautical and Aviation Engineering\\
	The Hong Kong Polytechnic University\\
	Hung Hom, Hong Kong SAR, China \\
	\texttt{lt.hsu@polyu.edu.hk} \\
	\And
	{\hspace{1mm}Yang~Jiang} \\
	Department of Geomatics Engineering\\
	University of Calgary\\
	Calgary T2N1N4, Canada \\
	\texttt{yang.jiang1@ucalgary.ca} \\
	\And
	\href{https://orcid.org/0000-0003-1797-2941}{\includegraphics[scale=0.06]{orcid.pdf}\hspace{1mm}Baoyu~Liu} \\
	Department of Geomatics Engineering\\
	University of Calgary\\
	Calgary T2N1N4, Canada \\
	\texttt{baoyu.liu@ucalgary.ca} \\
	\And
	\href{https://orcid.org/0000-0002-0565-2038}{\includegraphics[scale=0.06]{orcid.pdf}\hspace{1mm}Zhetao~Zhang} \\
	School of Earth Sciences and Engineering\\
	Hohai University\\
	Nanjing 211100, China \\
	\texttt{ztzhang@hhu.edu.cn} \\
	\And
	{\hspace{1mm}Yan~Xiang} \\
	School of Electronic Information \\and Electrical Engineering\\
	Shanghai Jiao Tong University\\
	Shanghai 200240, China \\
	\texttt{yan.xiang@sjtu.edu.cn} \\
	\And
	\href{https://orcid.org/0000-0001-6505-0832}{\includegraphics[scale=0.06]{orcid.pdf}\hspace{1mm}Naser~El-Sheimy} \\
	Department of Geomatics Engineering\\
	University of Calgary\\
	Calgary T2N1N4, Canada \\
	\texttt{elsheimy@ucalgary.ca} \\
}

\renewcommand{\shorttitle}{}

\begin{document}
\maketitle

\begin{abstract}
Many multi-sensor navigation systems urgently demand accurate positioning initialization from global navigation satellite systems (GNSSs) in challenging static scenarios. However, ground blockages against line-of-sight (LOS) signal reception make it difficult for GNSS users. Steering local codes in GNSS basebands is a desiring way to correct instantaneous signal phase misalignment, efficiently gathering useful signal power and increasing positioning accuracy. Besides, inertial navigation systems (INSs) have been used as a well-complementary dead reckoning (DR) sensor for GNSS receivers in kinematic scenarios resisting various interferences since early. But little work focuses on the case of whether the INS can improve GNSS receivers in static scenarios. Thus, this paper proposes an enhanced navigation system deeply integrated with low-cost INS solutions and GNSS high-accuracy carrier-based positioning. First, an absolute code phase is predicted from base station information, and integrated solution of the INS DR and real-time kinematic (RTK) results through an extended Kalman filter (EKF). Then, a numerically controlled oscillator (NCO) leverages the predicted code phase to improve the alignment between instantaneous local code phases and received ones. The proposed algorithm is realized in a vector-tracking GNSS software-defined radio (SDR). Real-world experiments demonstrate the proposed SDR regarding estimating time-of-arrival (TOA) and positioning accuracy. 
\end{abstract}


\section{Introduction}
Demands for GNSS devices will keep increasing over the following decades due to the explosion of smartphone-based navigation \cite{Sharma2020} and intelligent transportation construction \cite{Zhang2020}. Therefore, realizing accurate positioning in challenging environments using global navigation satellite system (GNSS) devices is a hot debate these days. However, current GNSS receiver techniques are uneasy to achieve a next-generation positioning, navigation, and timing (PNT) performance due to the intrinsic mechanism of GNSS electromagnetic waveforms. For instance, unlike LTE/5G wireless communication signals with an orthogonal frequency division multiple access (OFDMA) and substantial transmission power \cite{Liu2014a,Cimini1985}, the GNSS signals are transmitted at the same frequency. Differently, each signal channel is divided through the code division multiple access (CDMA). The GNSS signals are also confronted with severe channel fading over a long-distance transmission (approximately 20,000 km for the GNSS satellites operated in the medium Earth orbit). The former is naturally immune to the multipath effect, while the GNSS signals are less capable of resisting these interferences in the transmission. Due to this, recent research proposed a hybrid optical--wireless network that achieves a decimeter-level terrestrial positioning and sub-nanosecond timing aiming to enact as a supplement or even substitute the GNSS device in the future commercial market \cite{Koelemeij2022}. 

Against this background, there are many remaining problems and vast space regarding new GNSS receiver design, especially in challenging cases, and it is urgent to embark on renewing the current commercial receiver architecture. The CDMA signals are sensitive to the non-line-of-sight (NLOS) ray within one chip range causing big issues in channel estimation. In recent years, super-resolution algorithms (SRAs) emerged in GNSS signal processing to separate line-of-sight (LOS) and NLOS signals into different orthogonal spaces \cite{Krasner2022,Luo2021sra,DaRosaZanatta2020}. Recent work presented a graph Fourier transform (GFT) filter denoising the complex correlator outputs to replace the old GNSS tracking loop, which can be considered as a direct way to steer the code phase in challenging cases \cite{Luo2022}. Except for the separation using the state-of-the-art SRAs or the GNSS antenna changes \cite{Suzuki2020,Hong2020,Daneshmand2013}, modeling the superposition signals formed with LOS and NLOS rays is mainstream in the current GNSS community to overcome multipath interference \cite{Lau2007,Yan2022,Smolyakov2020}. 

Early in the 1990s, research revealed the prominence of GNSS products to overcome the predicament of tracking accurate Doppler frequency between the users' end and the satellite------vector delay lock loop (VDLL) \cite{Parkinson1996}. The vector approach is an intelligent choice to model the LOS Doppler frequency into a more proper shape. The basic idea of this technique is to leverage the user's navigation estimates to predict the Doppler frequency as compensation for the time of arrival (TOA) estimation in the GNSS baseband processing, stepping into a more accurate single-point positioning (SPP) solution. The VDLL allows GNSS researchers to optimize baseband signal modeling with information from multiple channels instead of a conservative loop filter algorithm in a single channel. Later, this idea was extended to assist the carrier phase modeling, for which it was nominated as a vector phase lock loop (VPLL) \cite{Zhodzishsky1998}. Improved versions and more specific experimental results based on the VPLL techniques have been presented over the following years \cite{Henkel2009,Shafaati2018a}. However, the stability and convergence of the VPLL are vulnerable to being destroyed when the biased error in the code phase modeling cannot be well removed (meaning that a multipath effect interferes with the GNSS baseband and its high-precision navigation solutions). It can be explained that the GNSS carrier and code signals are synchronized and significantly interact with each other. 

The wireless communication theory indicates that the fast fading \cite{Satyanarayana2012}, such as the multipath effect on carrier signals, causes a modeling error of a maximum of approximately one-fourth of a signal wavelength, e.g., about 5 cm for global positioning system (GPS) L1 signals \cite{Kelly2001}. This value is much smaller than the GNSS code phase error. For example, a typical value of the multipath effect on the code signal is commonly at the meter level, which is two orders of the magnitude of the carrier phase \cite{VANNEE1992}. Partially resorting to this mentioning, carrier-aiding is always used in a traditional GNSS baseband to support the code signal estimation \cite{Kaplan2017}. Similarly, a vector delay/frequency lock loop (VDFLL) was also presented to enhance the code phase tracking by combining the carrier-aiding and the VDLL approaches \cite{Lashley2009a}. 

The conventional vector tracking techniques impose an indirect approach to improving the TOA modeling over the code tracking process inside a GNSS receiver. The vector tracking has the potential to enable the baseband to yield a more accurate code Doppler frequency production. Then, the improved code Doppler replicates instantaneous code phases (i.e., TOA modeling) more precisely, contributing to a higher-quality navigation solution. This type of vector receiver can ultimately alleviate the harmful interference to the LOS signal estimation, especially when the user's end is moving \cite{DIetmayer2020}. Nevertheless, little efficiency will be available in the traditional vector tracking loop once the received LOS and NLOS rays cannot be well discriminated in terms of the Doppler frequency feature (i.e., the frequency or Fourier domain). Unfortunately, this case often occurs to a GNSS user today. 

The authors' recent research presented a method to optimize the TOA model in the GNSS tracking process assisted by an absolute position solution, not simply relying on the code frequency error between the baseband signal replica and the mapped code signal prediction from the user's navigation solution \cite{Luo2022aa}. More specifically, we take advantage of the high-accuracy positioning solution (more accurate than the code-based-only positioning result) to improve the TOA modeling aided by the vector tracking technique in the code phase domain instead of the code frequency domain. By coincidence, a recent paper using the classic SRA, i.e., root MUSIC, was also published aiming to achieve an analogous goal as stated before \cite{Krasner2022}. 

An inertial navigation system (INS) can resist the high-frequency random noise in the user's navigation as it works upon autonomous dead reckoning (DR), not relying on external information \cite{Groves2013}. Hence, integrating INS has been a prevalent way to rug the GNSS-based navigation system since early. To date, it has attracted much attention in the navigation field to what extent the low-cost INS increases the GNSS-based navigation \cite{Harke2022,Zhang2022}. However, research space remains in terms of the influence of the low-cost INS on GNSS baseband signal processing. 

Regarding these discussions, this work proposes an improved version of the authors' previous research \cite{Luo2022aa,Luo2021apa}. More specifically, a low-cost inertial measurement unit (IMU) is deeply integrated into a GNSS vector delay/frequency/phase lock loop (VDFPLL) software-defined radio (SDR). Meanwhile, an extended Kalman filter (EKF) is used to fuse float-RTK solutions and INS DR results. Hence, compared to our previous research \cite{Luo2022aa}, the main contributions of this work include the following:
\begin{enumerate}
\item  A low-cost IMU is combined with the float-RTK solutions via an EKF; integrated navigation solutions are used to improve the GNSS code phase domain TOA modeling directly; 
\item  An RTK/INS VDFPLL SDR is proposed and developed where the integration of RTK solutions and INS dead reckoning results, traditional scalar tracking loop (STL), VDFLL, and VDFPLL are realized and combined; 
\item  An approach showing how the INS enhances the GNSS baseband in a static scenario is presented based on real-world experiments, which few previous research discussed. 
\end{enumerate}

\begin{figure}[htbp]%
\centerline{\includegraphics[width=16cm]{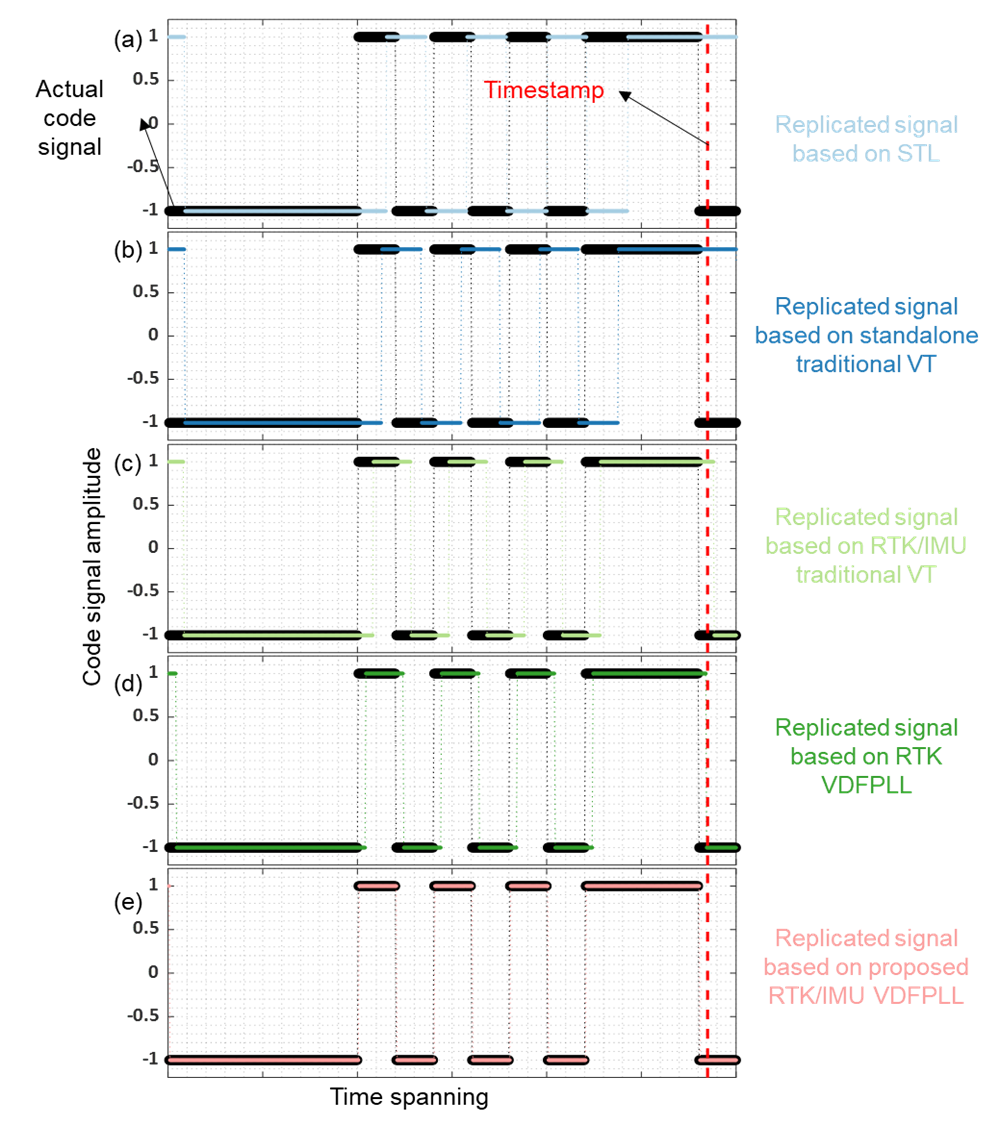}}%
\caption{Diagrammatic sketch reflecting the actual and locally replicated code signals varying with the time regarding different algorithms where colored curves correspond to the locally replicated code signals. (a) code phase misalignment is caused by the code frequency error and the initial (absolute) code phase error; (b) moderate frequency error is reduced (c) significant frequency error is reduced (d) significant frequency error and moderate initial code phase error are reduced (e) significant frequency error and significant initial code phase error are reduced.}%
\label{fig:fig1}%
\end{figure}

A diagram explaining the difference between the proposed and traditional algorithms towards the code phase estimation in the GNSS baseband is provided in \reffig{fig:fig1}. Before explaining \reffig{fig:fig1}, it is worthwhile to emphasize that the instantaneous code phase error consists of two primary parts regarding standard GNSS baseband processing. They are an absolute one from the initial code phase error and a relative one caused by the Doppler frequency error in which the received signal subtracts the local replica. These lead to the following discussions.

At first, the conventional STL causes an apparent code Doppler frequency error and initial code phase error (see \reffig{fig:fig1}(a)) \cite{Kaplan2017}; then, the traditional vector tracking compensates for parts of the Doppler frequency error reducing the relative code phase error  (see \reffig{fig:fig1}(b)) \cite{Parkinson1996,Lashley2021}; next, when the vector tracking technique is further aided with an IMU sensor, the GNSS baseband becomes more capable of alleviating the frequency error \cite{Lashley2013a}, but the initial code phase error is remained (see \reffig{fig:fig1}(c)); after that, when the RTK-based absolute-position-aided (APA) technique is involved in tracking, the initial code phase error can be reduced (see \reffig{fig:fig1}(d)) \cite{Luo2022aa}; finally, this work proposes a deep integration method of INS and GNSS RTK processing to correct a more absolute code phase error in the local replica (see \reffig{fig:fig1}(e)).

\begin{figure}[htbp]%
\centerline{\includegraphics[width=8.4cm]{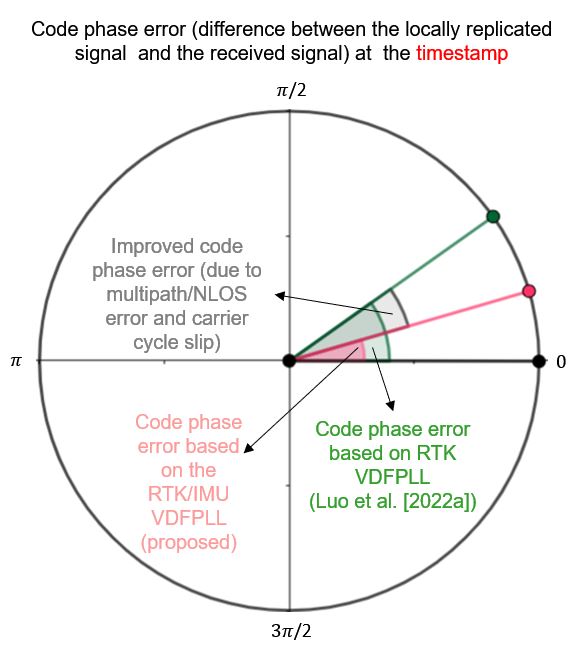}}
\caption{Comparison of the code phase error at the timestamp (see the dashed red line in \reffig{fig:fig4}) for extracting the instantaneous GNSS measurements (e.g., pseudoranges and carrier phases). }
\label{fig:fig2}
\end{figure}

\reffig{fig:fig2} further depicts the code phase errors at the timestamp (see the dashed red lines in \reffig{fig:fig1}) in the tracking process regarding the RTK-only APA and RTK/INS APA techniques. It is worth emphasizing that the timestamp denotes the local clock count to get the TOA estimation in the GNSS baseband (i.e., the time to extract the instantaneous GNSS measurements). Compared to our previous work \cite{Luo2022aa}, the proposed algorithm can improve the code phase estimation by removing the initial code phase error related to the multipath/NLOS effect and the carrier cycle slip (because of the involvement of the RTK-based APA technique). However, the case always occurs in the real world: for example, the multipath interference in a static GNSS user's receiver will cause such an absolute code phase error issue which is challenging in the current GNSS community. Therefore, this research comes up with a method to solve it. 

The remaining of this paper is organized as follows: Section \ref{sec2} introduces the methodology where the proposed VDFPLL based on RTK/INS deep integration is discussed in detail; before that, the RTK-position-aided VDFPLL is briefly introduced; then, two real-world stationary experiments are provided, and discussed in Section \ref{sec3}; finally, Section \ref{sec4} concludes this work.

\section{Materials and Methods}
\label{sec2}
This section will investigate how the APA code phase tracking in a GNSS baseband is realized with the proposed VDFPLL (deeply integrated with the float-RTK positioning and the INS DR). We will first recap the VDFPLL based on standalone GNSS RTK solutions; then, its improved form, deeply integrating the INS DR navigation solutions, will be discussed. Finally, how the proposed RTK/INS VDFPLL are combined with the STL and the VDFLL in a GPS SDR will be elaborated on.

\subsection{RTK-position-aided VDFPLL}
As mentioned earlier, the VDFPLL provides a way to directly steer the local code replica with the user's absolute position in the code phase domain instead of the conventional code frequency domain. Our previous work achieved this goal by presenting a practical means in the baseband that applies the user's RTK solution as a source of high-accuracy code phase prediction. This technique will be briefly stated in the following for the integrity of this work. 

\begin{figure}[htbp]%
\centerline{\includegraphics[width=12.8cm]{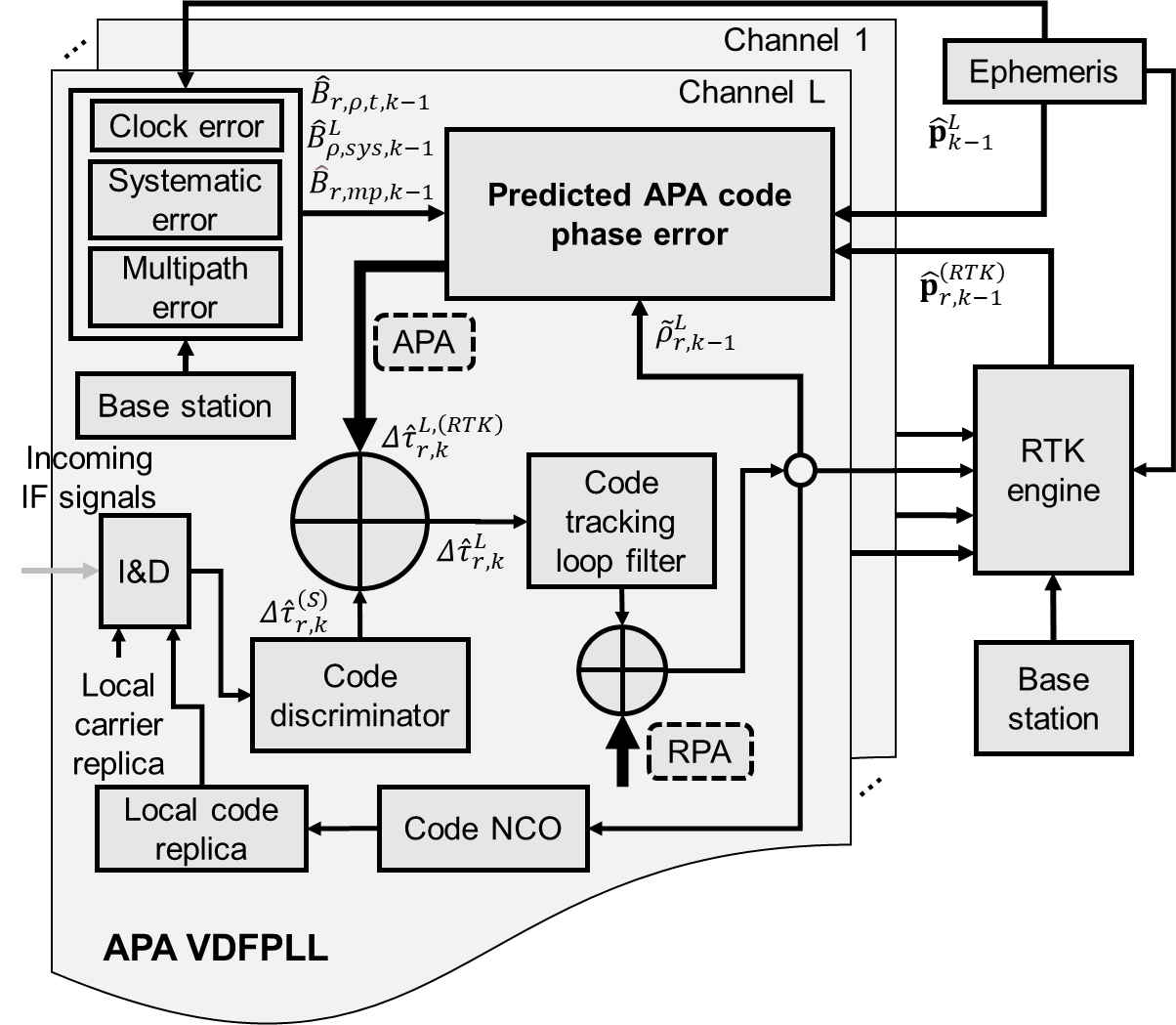}}
\caption{Overview of the GNSS baseband architecture with the RTK-position-aided VDFPLL \cite{Luo2022aa,Luo2021apa}.}
\label{fig:fig3}
\end{figure}

The architecture of the RTK-position-aided VDFPLL is illustrated in \reffig{fig:fig3}, where APA and RPA correspond to absolute- and relative-position-aided, respectively. It is worth noting that the RPA is achieved with the traditional VDFLL technique. Besides, the absolute code phase is also tracked aided by the vector tracking technique in the code phase domain. In this case, the following discriminates the entire code phase error 

\begin{equation}
\label{GrindEQ__1_}
\Delta {\hat{\tau }}^i_{r,k}=\Delta {\hat{\tau }}^{i,\left(S\right)}_{r,k}+\Delta {\hat{\tau }}^{i,\left(RTK\right)}_{r,k} 
\end{equation}
with
\begin{equation}
\label{GrindEQ__2_}
\Delta{\hat{\tau}}^{i,\left(RTK\right)}_{r,k}=\frac{f_c}{c}\left({\tilde{\rho}}^i_{r,k-1}-{\hat{\rho}}^{i,\left(RTK\right)}_{r,k-1}\right) 
\end{equation}

\begin{equation*}
{\hat{\rho}}^{i,\left(RTK\right)}_{r,k-1}={\hat{r}}^{i,\left(RTK\right)}_{r,k-1}+\left({\hat{B}}_{r,\rho,t,k-1}+{\hat{B}}^i_{\rho,sys,k-1}\right)-{\kappa }_D{\hat{B}}_{r,mp,k-1}
\end{equation*}
\begin{equation*}
{\hat{r}}^{i,\left(RTK\right)}_{r,k-1}=\left\|{\hat{\boldsymbol{\mathrm{p}}}}^i_{k-1}-{\hat{\boldsymbol{\mathrm{p}}}}^{\left(RTK\right)}_{r,k-1}\right\|
\end{equation*}

where subscript $k$ denotes the index of tracking epochs; $\Delta{\hat{\tau}}^{i,\left(S\right)}_{r,k}$ is the traditional discriminated code phase error through an early-minus-late-envelope code discriminator; $\Delta{\hat{\tau}}^{i,\left(RTK\right)}_{r,k}$ is the code phase error obtained from the APA approach; ${\tilde{\rho}}^i_{r,k-1}$ and ${\hat{\rho}}^{i,\left(RTK\right)}_{r,k-1}$ are the pseudoranges measured from the code tracking filter and predicted from the float RTK solution, respectively; ${\hat{r}}^{i,\left(RTK\right)}_{r,k-1}$ is the predicted geometry distance; ${\hat{\boldsymbol{\mathrm{p}}}}^i_{k-1}$ is the vector of satellite position; ${\hat{\boldsymbol{\mathrm{p}}}}^{\left(RTK\right)}_{r,k-1}$ is the vector of the estimated float RTK position; $\left({\hat{B}}_{r,\rho ,t,k-1}+{\hat{B}}^i_{\rho ,sys,k-1}\right)$ is the summation of the local clock bias error estimation and systematic error estimation, and it is computed from base station information and master satellite measurements \cite{Luo2022aa}; ${\hat{B}}_{r,mp,k-1}$ is the estimated multipath delay error imposed on the absolute code phase error via a between-satellite single difference algorithm, and ${\kappa }_D$ is its tuned coefficient constant based on the involved early-late spacing \cite{Luo2021apa}.

Next, the work process of the RTK-position-aided VDFPLL in the GPS SDR within the same tracking epoch is stated as follows:

\textbf{Step 1}: the SDR receives the incoming intermediate frequency (IF) GPS L1 C/A data via front-end equipment; 

\textbf{Step 2}: integration and dumping (I\&D) procedures upon correlators are implemented between the local code replica and the incoming IF GPS signals; 

\textbf{Step 3}: the correlator output passing through the traditional code discriminator yields $\mathrm{\Delta }{\hat{\tau }}^{\left(S\right)}_{r,k}$;  

\textbf{Step 4}: the bias of the discriminated code error compensated by the RTK-position aided (i.e., the APA operation) code error estimation $\Delta{\hat{\tau}}^{i,\left(RTK\right)}_{r,k}$ gives $\Delta{\hat{\tau}}^i_{r,k}$; 

\textbf{Step 5}: a code tracking loop filter denoises the code phase error from \textbf{Step 4};

\textbf{Step 6}: an RPA technique (i.e., the VDFLL) is executed to alleviate the code frequency error in the code-tracking process; 

\textbf{Step 7}: a numerically controlled oscillator (NCO) leverages the output of \textbf{Step 6} to produce the TOA estimation (the raw output of the code loop filter aided by the Doppler prediction), the pseudorange measurements (be de-noised by the carrier smoothing technique);

\textbf{Step 8}: the RTK engine leverages the pseudoranges and carrier phases (the raw output of the carrier loop filter aided by the Doppler prediction) from all the tracking channels, navigation data, and the base station information to compute the float RTK solutions; 

\textbf{Step 9}: the APA code phase error $\Delta{\hat{\tau}}^{i,\left(RTK\right)}_{r,k}$ is computed with the float-RTK solutions and the pseudorange measurement by \eqref{GrindEQ__2_}; 

\textbf{Step 10}: repeating \textbf{Step 2}, the RTK-position-aided VDFPLL is working for the next tracking epoch. 

To conclude, the work process of the VDFPLL based on the float-RTK solutions executed in a GNSS SDR has been browsed.

\subsection{The proposed VDFPLL based on RTK/INS deep integration}

\subsubsection{Architectures of the proposed VDFPLL SDR}
\begin{figure}[htbp]%
\centering
\subfloat{\centerline{\includegraphics[width=10cm]{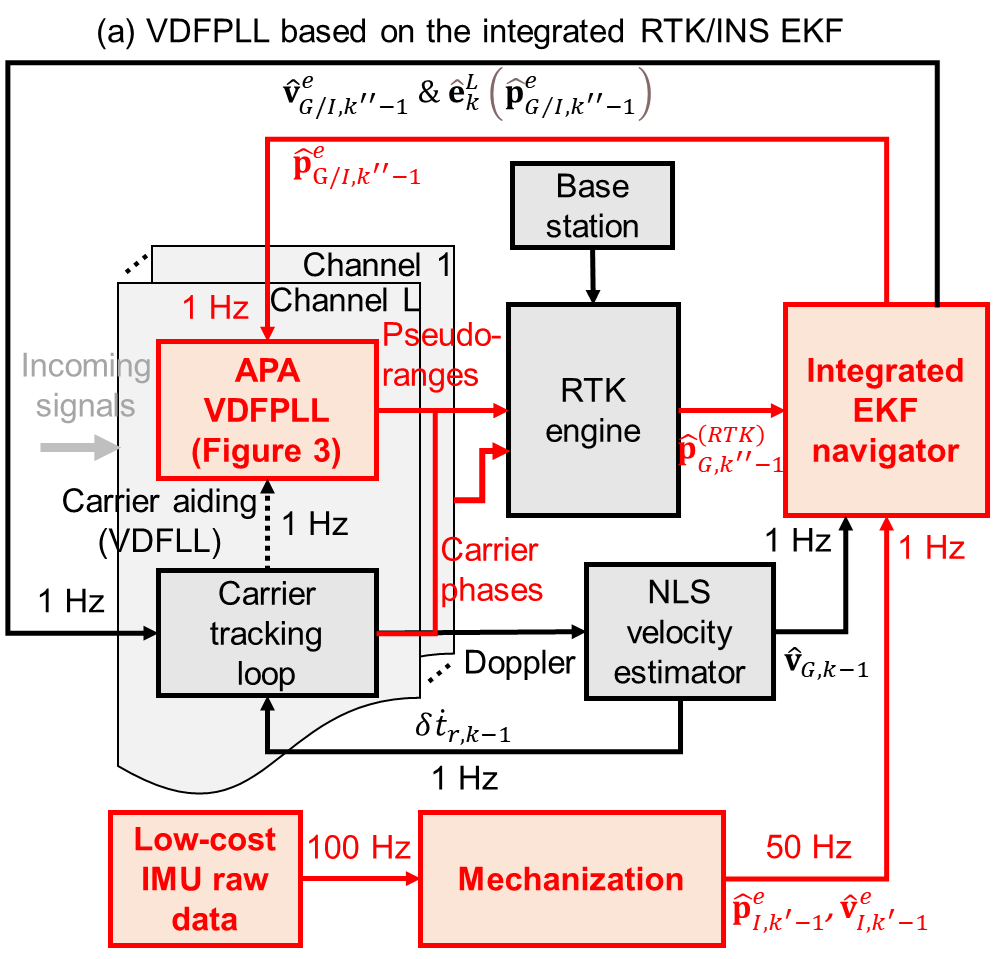}}%
\label{fig:fig4a}}
\hfil
\subfloat{\includegraphics[width=8cm]{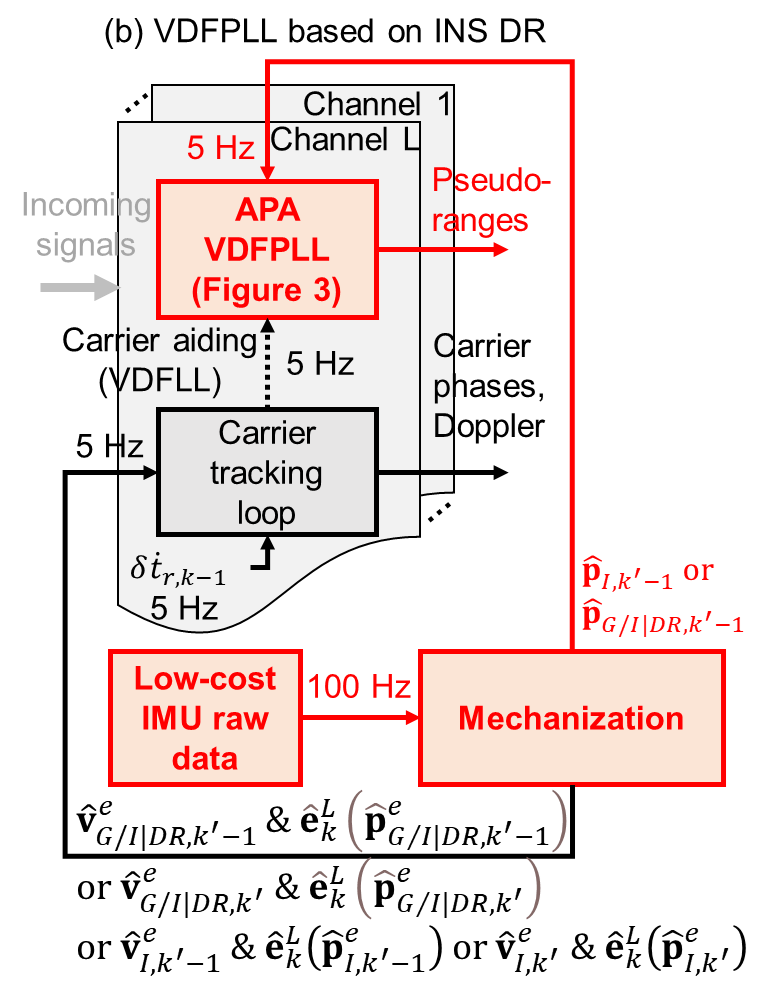}%
\label{fig:fig4b}}
\hfil
\subfloat{\includegraphics[width=7cm]{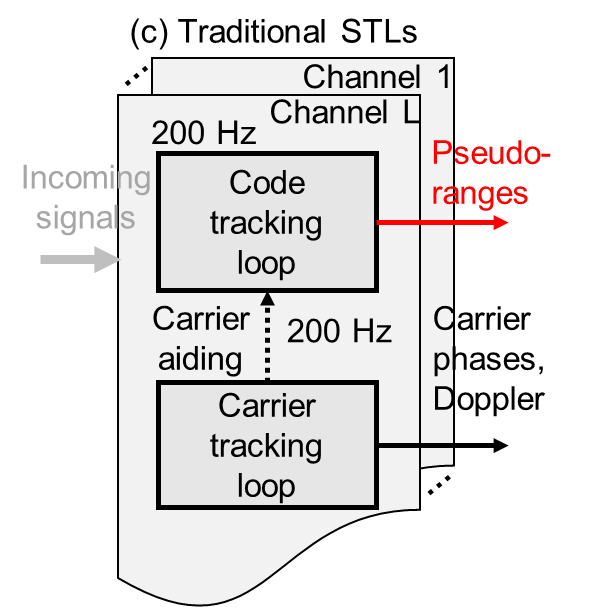}%
\label{fig:fig4c}}
\caption{Architectures of the proposed VDFPLL-enhanced GPS SDR based on the deep integration of float RTK solutions and INS DR navigation results (detailed discussions refer to the \textbf{Algorithm 1} stated later.)}%
\label{fig:fig4}%
\end{figure}

The architectures of the proposed APA VDFPLL GPS SDR deeply integrated with the float RTK solutions and INS dead reckoning results are displayed in \reffig{fig:fig4}. It is worthwhile to mention that hybrid tracking loops are adopted here due to the data rates discrepancy corresponded to various sources, i.e., the proposed SDR tracking, the base station, and the IMU sensor raw data.

\noindent 

First, there are two procedures for updating the code tracking loop with the APA method in the SDR. On the one hand, the TOA model is predicted from the integrated float RTK/INS EKF, as depicted in \reffig{fig:fig4}(a). On the other hand, the proposed VDFPLL updating rate is 5 Hz, that is higher than the RTK solution rate, so the INS DR navigation solutions (with a rate of 50 Hz) are interpolated in the updating process when the RTK solutions (with a rate of 1 Hz) are absent. It is worth noting that we take two samples of the IMU raw data per update to compensate for coning and sculling errors, so the raw data rate of the IMU is 100 Hz, while the DR navigation results are at the rate of 50 Hz. 

Finally, as the tracking loop updating rate in the proposed SDR baseband is 200 Hz, much higher than the VDFPLL rate, the traditional STLs for code and carrier tracking are interpolated across the intervals where the VDFPLL is not activated. 

As a result, the three tracking loops are jointly working in the proposed RTK/INS-based VDFPLL SDR, and they are the VDFPLL based on the RTK/INS integrated EKF (see \reffig{fig:fig4}(a)), the VDFPLL based on the INS DR (see \reffig{fig:fig4}(b)), and the traditional STL (see \reffig{fig:fig4}(c)).

\subsubsection{RTK/INS EKF Navigator and INS DR }
As illustrated in \reffig{fig:fig4}, there are three types of TOA estimation formed in the proposed GNSS SDR corresponding to Figures 4(a), 4(b), and 4(c), respectively. They will be elaborated on one by one in the following.

At first, one is formed using the absolute position estimated from the integrated RTK/INS EKF navigator (see \reffig{fig:fig4}(a)). The RTK engine comes from an open-source package goGPS v0.4.3 \cite{Herrera2016}. Then, the float RTK deeply integrated into the proposed SDR has been introduced in the authors' previous work \cite{Luo2022aa}. Then, the SDR platform to realize the deep integration of RTK and INS has also been built and investigated in the authors' previous publications \cite{Luo2019,Luo2021utc}. 

Besides, we will discuss the EKF algorithm used in this work. The state transition equation is given by
\[\delta{\boldsymbol{x}}^e_{k''}=\bf{\Phi }^e_{k'',k''-1}\delta{\boldsymbol{x}}^e_{k''-1}+{\boldsymbol{w}}^e_{k''-1}\] 
with
\[k''=\left\lfloor k^{\mathrm{'}}\mathrm{/}K\right\rfloor \] 
\[k'=\left\lfloor k/M\right\rfloor \] 
where superscript $e$ represents the Earth-centered, Earth-fixed (ECEF) coordinate frame; subscript $k''$ denotes the epoch index of the EKF updates; $K$ is the integer ratio of the INS DR and the EKF updating rates, where the former and the latter are 50 Hz and 1 Hz (constrained by the rate of the base station information), respectively; $k'$ is the epoch index of INS DR solutions; $M$ is the integer ratio of the GNSS tracking rate and the INS updating rate where the tracking rate (200 Hz) is no lower than the INS updating rate (50 Hz) here; so it satisfies $K,M\in {\mathbb{Z}}^+$; $\bf{\Phi }^e_{k'',k''-1}$ is the transition matrix; ${\boldsymbol{w}}^e_{k''-1}$ is the process noise vector. Then, the state vector of the EKF model in the ECEF frame is given by 
\[\delta{\boldsymbol{x}}^e_{k''}={\left[{\left(\delta\boldsymbol{\mathrm \psi }^e\right)}^T,{\left(\delta{\boldsymbol{\mathrm {v}}}^e\right)}^T,{\left(\delta{\boldsymbol{\mathrm{p}}}^e\right)}^T,{\left(\delta{\boldsymbol{b}}_g\right)}^T,{\left(\delta{\boldsymbol{b}}_a\right)}^T\right]}^T_{k''}\] 
where $\delta\boldsymbol{\psi }^e$ is the attitude error vector; $\delta{\boldsymbol{\mathrm v}}^e$ is the 3D velocity error vector; $\delta{\boldsymbol{\mathrm p}}^e$ is the 3D position error vector; $\delta{\boldsymbol{b}}_g$ and $\delta{\boldsymbol{b}}_a$ are the respective gyro and accelerometer bias error vectors. 

Next, the observation equation is provided as 
\[\delta{\boldsymbol{z}}_{k''}={\boldsymbol{H}}^e_{k''}\delta{\boldsymbol{x}}^e_{k''}+{\boldsymbol{v}}_{k''}\] 
where ${\boldsymbol{H}}^e_{k''}$ is the observation matrix; ${\boldsymbol{v}}_{k''}$ is the observation noise vector. The observation vector, including position errors and velocity errors, is provided as
\begin{equation*}
\delta {{\bf{z}}_{k''}} = {\left[ {{{\left[ {\begin{array}{*{20}{c}}
{{\tilde{\dot x}_I} - {\tilde {\dot x}_{G,NLS}}}\\
{{\tilde {\dot y}_I} - {\tilde {\dot y}_{G,NLS}}}\\
{{\tilde {\dot z}_I} - {\tilde {\dot z}_{G,NLS}}}
\end{array}} \right]}^T},{{\left[ {\begin{array}{*{20}{c}}
{{\tilde {x}_I} - {{\tilde x}_{G,RTK}}}\\
{{\tilde {y}_I} - {{\tilde y}_{G,RTK}}}\\
{{\tilde {z}_I} - {{\tilde z}_{G,RTK}}}
\end{array}} \right]}^T}} \right]_{k''}}^T
\end{equation*}
where $\left[\tilde{x},\tilde{y},\tilde{z}\right]$ and $\left[\tilde{\dot{x}},\tilde{\dot{y}},\tilde{\dot{z}}\right]$ correspond to the 3D positions and velocities w.r.t. the ECEF frame, respectively; subscript $I$ and $G$ correspond to the solutions obtained from the INS and the GNSS, respectively; the subscript ``RTK'' means that the GNSS position results are solved by the float RTK algorithm \cite{Luo2022aa}, and the subscript ``NLS'' represents that the GNSS velocity results are calculated from the standard non-linear squared (NLS) method \cite{Luo2019b}. How to build $\bf{\Phi }^e_{k''\ ,k''-1}$ and ${\boldsymbol{H}}^e_{k''}$, as well as how to form the process noise covariance matrix and the observation noise covariance matrix can refer to \cite{Luo2019}. 

After the system model is built, the recursive estimation of the EKF algorithm is to predict and update the state vector \cite{Faragher2012}. Finally, upon the time epoch where base station information is available for the RTK algorithm, the navigation solutions corresponding to the velocity, position, and attitude information from the EKF algorithm are given by
\begin{equation*}
{\hat{\boldsymbol{\mathrm{C}}}}^e_{G/I,b,k''-1}\boldsymbol{\approx }\left({\boldsymbol{\mathrm{I}}}_3-\left(\delta{\hat{\boldsymbol\psi}}^e_{k''-1}\right)\times \right){\hat{\boldsymbol{\mathrm{C}}}}^e_{I,b,k'-1}
\end{equation*}
\begin{equation*}
{\hat{\boldsymbol{\mathrm{v}}}}^e_{G/I,k''-1}={\hat{\boldsymbol{\mathrm{v}}}}^e_{I,k'-1}\boldsymbol{-}\delta {\hat{\boldsymbol{\mathrm{v}}}}^e_{k''-1}
\end{equation*}
\begin{equation*}  
{\hat{\boldsymbol{\mathrm{p}}}}^e_{G/I,k''-1}\boldsymbol{=}{\hat{\boldsymbol{\mathrm{p}}}}^e_{I,k'-1}\boldsymbol{-}\delta {\hat{\boldsymbol{\mathrm{p}}}}^e_{k''-1} 
\end{equation*} 
where $\left(\cdot \right)\times $ denotes the skew matrix operator; ${\boldsymbol{\mathrm{I}}}_3$ is the 3-order identity matrix; $\delta {\hat{\boldsymbol{\mathrm{v}}}}^e_{k''-1}$\textbf{ }$\delta {\hat{\boldsymbol{\mathrm{p}}}}^e_{k''-1}$, and $\delta\hat{\boldsymbol{\psi }}^e_{k''-1}$ are the estimated state vectors about velocity and position errors, and attitude errors, and ${\hat{\boldsymbol{\mathrm{v}}}}^e_{G/I,k''-1}$\textbf{, }${\hat{\boldsymbol{\mathrm{p}}}}^e_{G/I,k''-1}$\textbf{ }and ${\hat{\boldsymbol{\mathrm{C}}}}^e_{G/I,k''-1}$\textbf{ }are the estimated navigation vectors (corresponding to the respective velocity, position, and attitude); ${\hat{\boldsymbol{\mathrm{v}}}}^e_{I,k'-1}$\textbf{, }${\hat{\boldsymbol{\mathrm{p}}}}^e_{I,k'-1}$\textbf{ }and ${\hat{\boldsymbol{\mathrm{C}}}}^e_{I,b,k'-1}$\textbf{ }are the counterparts solely upon the INS DR process, which will be introduced subsequently. 

Within the epochs where the base station information is missing, the EKF-based results at the previous epoch can contribute to the INS DR process at the current epoch as the following
\[{\hat{\boldsymbol{\mathrm{v}}}}^e_{G/I|DR,k'}={\hat{\boldsymbol{\mathrm{v}}}}^e_{G/I,k''-1}\boldsymbol{+}\left({\overline{\boldsymbol{\mathrm{C}}}}^e_{G/I,k''-1}{\tilde{\boldsymbol{\mathrm{f}}}}^b_{ib,k'-1}\boldsymbol{-}2\left({\boldsymbol{\omega }}_{ie}\right)\times {\hat{\boldsymbol{\mathrm{v}}}}^e_{G/I,k''-1}+{\boldsymbol{\mathrm{g}}}^e\left({\hat{\boldsymbol{\mathrm{p}}}}^e_{G/I,k''-1}\right)\right)\mathrm{\Delta }t_I\] 
\[{\hat{\boldsymbol{\mathrm{p}}}}^e_{G/I|DR,k'}\boldsymbol{=}{\hat{\boldsymbol{\mathrm{p}}}}^e_{G/I,k''-1}+\left({\hat{\boldsymbol{\mathrm{v}}}}^e_{G/I|DR,k'}\boldsymbol{+}{\hat{\boldsymbol{\mathrm{v}}}}^e_{G/I,k''-1}\right)\frac{\mathrm{\Delta }t_I}{2}\] 
\[{\hat{\boldsymbol{\mathrm{C}}}}^e_{G/I|DR,b,k'}\boldsymbol{=}{\boldsymbol{\mathrm{C}}}^e_i\left(\mathrm{\Delta }t_I\right){\hat{\boldsymbol{\mathrm{C}}}}^e_{G/I,b,k''-1}\left({\boldsymbol{\mathrm{I}}}_3+\left({\tilde{\boldsymbol{\omega }}}^b_{ib,k'-1}\mathrm{\Delta }t_I\right)\times \right)\] 
with
\[{\boldsymbol{\omega }}_{ie}={\left[ \begin{array}{ccc}
0 & 0 & {\omega }_{ie} \end{array}
\right]}^T\] 
\[{\boldsymbol{\mathrm{C}}}^e_i\left(\mathrm{\Delta }t_I\right)\boldsymbol{=}\left[ \begin{array}{ccc}
{\mathrm{cos} \left({\omega }_{ie}\mathrm{\Delta }t_I\right)\ } & {\mathrm{sin} \left({\omega }_{ie}\mathrm{\Delta }t_I\right)\ } & 0 \\ 
{\mathrm{-sin} \left({\omega }_{ie}\mathrm{\Delta }t_I\right)\ } & {\mathrm{cos} \left({\omega }_{ie}\mathrm{\Delta }t_I\right)\ } & 0 \\ 
0 & 0 & 1 \end{array}
\right]\] 
where $\mathrm{\Delta }t_I$ is the updating interval of the EKF; ${\tilde{\boldsymbol{\mathrm{f}}}}^b_{ib,k'-1}$ and ${\tilde{\boldsymbol{\omega }}}^b_{ib,k'-1}$ are the specific force and angular rate measurement vectors of the body frame w.r.t. ECEF frame; ${\omega }_{ie}$ is the Earth rotation rate, i.e., 7.292115e-5 rad/s and ${\boldsymbol{\omega }}_{ie}$ is its vector form; ${\boldsymbol{\mathrm{g}}}^e\left({\hat{\boldsymbol{\mathrm{p}}}}^e_{G/I,k''-1}\right)$ is the gravity acceleration vector function in the ECEF frame varying with the user's position ${\hat{\boldsymbol{\mathrm{p}}}}^e_{G/I\ ,k''-1}$ (see Equations (2.133) and (2.142) in \cite{Groves2013}); ${\boldsymbol{\mathrm{C}}}^e_i\left(\mathrm{\Delta }t_I\right)$\textbf{ }is the Earth rotation matrix from the Earth-centered inertial (ECI) to the ECEF coordinate frame varying with the updating interval $\mathrm{\Delta }t_I$; ${\overline{\boldsymbol{\mathrm{C}}}}^e_{G/I,b,k''-1}$ is\textbf{ }the averaging transformation matrix w.r.t. the body-to-ECEF-frame coordinate obtained from ${\hat{\boldsymbol{\mathrm{C}}}}^e_{G/I,b,k''-1}$\textbf{ }(see Equations (5.84) and (5.85) in \cite{Groves2013}). 

Then, considering the case where the navigating solutions are derived from the INS DR process (see \reffig{fig:fig4}(b)), the mechanization in the ECEF frame can be expressed as
\[{\hat{\boldsymbol{\mathrm{v}}}}^e_{I,k'}={\hat{\boldsymbol{\mathrm{v}}}}^e_{G/I|DR,k'-1}\boldsymbol{+}\left({\overline{\boldsymbol{\mathrm{C}}}}^e_{I,b,k'-1}{\tilde{\boldsymbol{\mathrm{f}}}}^b_{ib,k'-1}\boldsymbol{-}2\left({\boldsymbol{\omega }}_{ie}\right)\times {\hat{\boldsymbol{\mathrm{v}}}}^e_{G/I|DR,k'-1}+{\boldsymbol{\mathrm{g}}}^e\left({\hat{\boldsymbol{\mathrm{p}}}}^e_{G/I|DR,k'-1}\right)\right)\mathrm{\Delta }t_I\] 
\[{\hat{\boldsymbol{\mathrm{p}}}}^e_{I,k'}\boldsymbol{=}{\hat{\boldsymbol{\mathrm{p}}}}^e_{G/I|DR,k'-1}+\left({\hat{\boldsymbol{\mathrm{v}}}}^e_{I,k'}\boldsymbol{+}{\hat{\boldsymbol{\mathrm{v}}}}^e_{G/I|DR,k'-1}\right)\frac{\mathrm{\Delta }t_I}{2}\] 
\[{\hat{\boldsymbol{\mathrm{C}}}}^e_{I,b,k'}\boldsymbol{=}{\boldsymbol{\mathrm{C}}}^e_i\left(\mathrm{\Delta }t_I\right){\hat{\boldsymbol{\mathrm{C}}}}^e_{G/I|DR,b,k'-1}\left({\boldsymbol{\mathrm{I}}}_3+\left({\tilde{\boldsymbol{\omega }}}^b_{ib,k'-1}\mathrm{\Delta }t_I\right)\times \right)\] 
or
\[{\hat{\boldsymbol{\mathrm{v}}}}^e_{I,k'}={\hat{\boldsymbol{\mathrm{v}}}}^e_{I,k'-1}\boldsymbol{+}\left({\overline{\boldsymbol{\mathrm{C}}}}^e_{I,b,k'-1}{\tilde{\boldsymbol{\mathrm{f}}}}^b_{ib,k'-1}\boldsymbol{-}2\left({\boldsymbol{\omega }}_{ie}\right)\times {\hat{\boldsymbol{\mathrm{v}}}}^e_{I,k'-1}+{\boldsymbol{\mathrm{g}}}^e\left({\hat{\boldsymbol{\mathrm{p}}}}^e_{I,k'-1}\right)\right)\mathrm{\Delta }t_I\] 
\[{\hat{\boldsymbol{\mathrm{p}}}}^e_{I,k'}\boldsymbol{=}{\hat{\boldsymbol{\mathrm{p}}}}^e_{I,k'-1}+\left({\hat{\boldsymbol{\mathrm{v}}}}^e_{I,k'}\boldsymbol{+}{\hat{\boldsymbol{\mathrm{v}}}}^e_{I,k'-1}\right)\frac{\mathrm{\Delta }t_I}{2}\] 
\[{\hat{\boldsymbol{\mathrm{C}}}}^e_{I,b,k'}\boldsymbol{=}{\boldsymbol{\mathrm{C}}}^e_i\left(\mathrm{\Delta }t_I\right){\hat{\boldsymbol{\mathrm{C}}}}^e_{I,b,k'-1}\left({\boldsymbol{\mathrm{I}}}_3+\left({\tilde{\boldsymbol{\omega }}}^b_{ib,k'-1}\mathrm{\Delta }t_I\right)\times \right)\] 
where ${\overline{\boldsymbol{\mathrm{C}}}}^e_{I,b,k'-1}$ is the averaging transformation matrix computed from ${\hat{\boldsymbol{\mathrm{C}}}}^e_{I,b,k'-1}$. 

It is worth mentioning that the tracking rate (200 Hz) is higher than the INS DR updating rate (50 Hz). So, three out of four tracking intervals do not have an update for the INS DR. Assuming that the user's navigation results are not changed significantly over the time 0.02s (i.e., $\frac{1}{50\mathrm{Hz}}$), when $k'=k/M$, we make an approximation that the navigation estimations in the following $\left(k+1\right)$th, $(k+2)$th, and $\left(k+3\right)$th tracking epochs are identical to the ones computed at the $k$th, to interpolate the tracking epochs without the INS updating.

\subsubsection{RTK/INS APA Code Phase Tracking}
The baseband TOA modeling at the start of the $k$th epoch aided by the absolute positions from the integrated RTK/INS EKF and the INS DR is estimated through

\begin{equation}\label{GrindEQ__3_}
{\widehat{TOA}}^i_k=c^{-1}{\tilde{\rho }}^i_k-\delta {\hat{t}}_{r,k}
\end{equation}

with the pseudorange model of 

\begin{equation}\label{GrindEQ__4_}
{\tilde{\rho }}^i_k\triangleq {\tilde{\rho }}^i_{k-1}+cf^{-1}_c\left(\left(f_c+{\hat{f}}^i_{code,dop,k}\right)T_{coh}+\mathrm{\Delta }{\hat{\tau }}^{i+,\left(RTK/INS\right)}_{code,k,0}\right)
\end{equation}

where ${\tilde{\rho }}^i_{k-1}$ and ${\tilde{\rho }}^i_k$ are the instantaneous pseudorange measurements at the respective previous and current epochs; $\delta {\hat{t}}_{r,k}$ is the estimated local clock bias error; $f_c$ and $c$ are the spreading code rate and the speed of light, respectively; $T_{coh}$ is the coherent integration time ${\hat{f}}^i_{code,dop,k}$ is the estimated code Doppler frequency and ${\mathit{\Delta}\hat{\tau }}^{i+,\left(RTK/INS\right)}_{code,k,0}$ is the proposed initial code phase error estimate in chips at the start of the $k$th epoch, and the estimation processes of them will be subsequently elaborated. 

On the one hand, ${\hat{f}}^i_{code,dop,k}$ can be written as 

\begin{equation}\label{GrindEQ__5_}
{\hat{f}}^i_{code,dop,k}=-\frac{f_c}{f_r}{\tilde{f}}_{carr,dop,k}+T^{-1}_{coh}\mathrm{\Delta }{\hat{\tau }}^{i+}_{r,k}
\end{equation}
\begin{equation} \label{GrindEQ__6_}
{\tilde{f}}^i_{carr,dop,k}=\mathrm{\Delta }{\hat{f}}^{i,\left(aid\right)}_{carr,k}+T^{-1}_{coh}\mathrm{\Delta }{\hat{\varphi }}^{i+}_{r,k}
\end{equation}

with

\begin{equation}\label{GrindEQ__7_}
\mathrm{\Delta }{\hat{f}}^{i,\left(aid\right)}_{carr,k}\mathrm{=}\frac{f_r}{c}\left({\hat{\boldsymbol{\mathrm{v}}}}^e_{G/I,k''-1}\cdot {\hat{\boldsymbol{\mathrm{e}}}}^i_k\left({\hat{\boldsymbol{\mathrm{p}}}}^e_{G/I,k''-1}\right)-{\hat{\boldsymbol{\mathrm{v}}}}^i_k\cdot {\hat{\boldsymbol{\mathrm{e}}}}^i_k\left({\hat{\boldsymbol{\mathrm{p}}}}^e_{G/I,k''-1}\right)+c\delta {\hat{\dot{t}}}_{r,k}-c\delta {\hat{\dot{t}}}^i_k\right)
\end{equation}
or 
\begin{equation}\label{GrindEQ__8_}
\mathrm{\Delta }{\hat{f}}^{i,\left(aid\right)}_{carr,k}\mathrm{=}\frac{f_r}{c}\left({\hat{\boldsymbol{\mathrm{v}}}}^e_{G/I|DR,k'-1}\cdot {\hat{\boldsymbol{\mathrm{e}}}}^i_k\left({\hat{\boldsymbol{\mathrm{p}}}}^e_{G/I|DR,k'-1}\right)-{\hat{\boldsymbol{\mathrm{v}}}}^i_k\cdot {\hat{\boldsymbol{\mathrm{e}}}}^i_k\left({\hat{\boldsymbol{\mathrm{p}}}}^e_{G/I|DR,k'-1}\right)+c\delta {\hat{\dot{t}}}_{r,k}-c\delta {\hat{\dot{t}}}^i_k\right)
\end{equation}
\begin{equation} \label{GrindEQ__9_}
\mathrm{\Delta }{\hat{f}}^{i,\left(aid\right)}_{carr,k}\mathrm{=}\frac{f_r}{c}\left({\hat{\boldsymbol{\mathrm{v}}}}^e_{G/I|DR,k'}\cdot {\hat{\boldsymbol{\mathrm{e}}}}^i_k\left({\hat{\boldsymbol{\mathrm{p}}}}^e_{G/I|DR,k'}\right)-{\hat{\boldsymbol{\mathrm{v}}}}^i_k\cdot {\hat{\boldsymbol{\mathrm{e}}}}^i_k\left({\hat{\boldsymbol{\mathrm{p}}}}^e_{G/I|DR,k'}\right)+c\delta {\hat{\dot{t}}}_{r,k}-c\delta {\hat{\dot{t}}}^i_k\right)
\end{equation}
\begin{equation}\label{GrindEQ__10_}
\mathrm{\Delta }{\hat{f}}^{i,\left(aid\right)}_{carr,k}\mathrm{=}\frac{f_r}{c}\left({\hat{\boldsymbol{\mathrm{v}}}}^e_{I,k'}\cdot {\hat{\boldsymbol{\mathrm{e}}}}^i_k\left({\hat{\boldsymbol{\mathrm{p}}}}^e_{I,k'}\right)-{\hat{\boldsymbol{\mathrm{v}}}}^i_k\cdot {\hat{\boldsymbol{\mathrm{e}}}}^i_k\left({\hat{\boldsymbol{\mathrm{p}}}}^e_{I,k'}\right)+c\delta {\hat{\dot{t}}}_{r,k}-c\delta {\hat{\dot{t}}}^i_k\right)
\end{equation}

\begin{equation}\label{GrindEQ__11_}
\mathrm{\Delta }{\hat{f}}^{i,\left(aid\right)}_{carr,k}\mathrm{=}\frac{f_r}{c}\left({\hat{\boldsymbol{\mathrm{v}}}}^e_{I,k'-1}\cdot {\hat{\boldsymbol{\mathrm{e}}}}^i_k\left({\hat{\boldsymbol{\mathrm{p}}}}^e_{I,k'-1}\right)-{\hat{\boldsymbol{\mathrm{v}}}}^i_k\cdot {\hat{\boldsymbol{\mathrm{e}}}}^i_k\left({\hat{\boldsymbol{\mathrm{p}}}}^e_{I,k'-1}\right)+c\delta {\hat{\dot{t}}}_{r,k}-c\delta {\hat{\dot{t}}}^i_k\right)
\end{equation}
where ${\tilde{f}}_{carr,dop,k}$ denotes the carrier Doppler frequency measurement; $\left(T^{-1}_{coh}\mathrm{\Delta }{\hat{\tau }}^{i+}_{r,k}\right)$ and $\left(T^{-1}_{coh}\mathrm{\Delta }{\hat{\varphi }}^{i+}_{r,k}\right)$ are the filtered code phase error and the filtered carrier phase error through the loop filters, respectively, which have accounted for the coherent integration interval in tracking, and its input is $\mathrm{\Delta }{\hat{\tau }}^i_{x,k}$ which will be explained later, with $x\in \left\{I,G/I,G/I|DR\right\}$; $\mathrm{\Delta }{\hat{f}}^{i,\left(aid\right)}_{carr,k}$ is the aided Doppler frequency computed via the user's velocity estimation, known as a VDFLL technique \cite{Lashley2021}; ${\hat{\boldsymbol{\mathrm{v}}}}_{x,k'}$ / ${\hat{\boldsymbol{\mathrm{v}}}}_{x,k''}$ and $\delta {\hat{\dot{t}}}_{r,k}$ are the predicted user's velocity vector and the predicted user's clock drift; ${\hat{\boldsymbol{\mathrm{v}}}}^i_k$ and $\delta {\hat{\dot{t}}}^i_k$ are the satellite velocity vector and the satellite clock drift predicted with the broadcast ephemeris; ${\hat{\boldsymbol{\mathrm{e}}}}^i_k\left(\cdot \right)$ is the operator of the unit cosine vector varied with the position estimation. 

As mentioned above, $\mathrm{\Delta }{\hat{\tau }}^i_{x,k}$ is the APA discriminated code phase error, and there are three ways to obtain this estimate in the code tracking loop. For instance, the ones estimated via the respective RTK/INS EKF solution, the two-consecutive-epoch INS DR, and INS DR right after the EKF are computed as 

\begin{equation}\label{GrindEQ__12_}
\mathrm{\Delta }{\hat{\tau }}^i_{G/I,k}=\mathrm{\Delta }{\hat{\tau }}^{i,\left(S\right)}_{r,k}+\mathrm{\Delta }{\hat{\tau }}^{i,\left(APA\right)}_{r,k}\left({\hat{r}}^i_{G/I,k-1}\right)
\end{equation}
\begin{equation}\label{GrindEQ__13_}
\mathrm{\Delta }{\hat{\tau }}^i_{I,k}=\mathrm{\Delta }{\hat{\tau }}^{i,\left(S\right)}_{r,k}+\mathrm{\Delta }{\hat{\tau }}^{i,\left(APA\right)}_{r,k}\left({\hat{r}}^i_{I,k-1}\right)
\end{equation}
\begin{equation}\label{GrindEQ__14_}
\mathrm{\Delta }{\hat{\tau }}^i_{G/I|DR,k}=\mathrm{\Delta }{\hat{\tau }}^{i,\left(S\right)}_{r,k}+\mathrm{\Delta }{\hat{\tau }}^{i,\left(APA\right)}_{r,k}\left({\hat{r}}^i_{G/I|DR,k-1}\right)
\end{equation}
with
\[\hat r_{G/I,k - 1}^i = \left\| {{\hat{\bf{ p}}}_{k - {\rm{1}}}^i - {\hat{\bf{p}}}_{G/I,k'' - 1}^e} \right\|\]
\[{\hat{r}}^i_{I,k-1}=\left\|{\hat{\boldsymbol{\mathrm{p}}}}^i_{k\boldsymbol{-}\mathrm{1}}-{\hat{\boldsymbol{\mathrm{p}}}}^e_{I,k'-1}\right\|\] 
\[{\hat{r}}^i_{G/I|DR,k-1}=\left\|{\hat{\boldsymbol{\mathrm{p}}}}^i_{k\boldsymbol{-}\mathrm{1}}-{\hat{\boldsymbol{\mathrm{p}}}}^e_{G/I|DR,k'-1}\right\|\] 
where $\mathrm{\Delta }{\hat{\tau }}^{i,(S)}_{r,k}$ is the traditional discriminated code phase error as introduced earlier; ${\hat{\boldsymbol{\mathrm{p}}}}^i_{k\boldsymbol{-}\mathrm{1}}$ is the satellite position vector computed from the broadcast ephemeris; $\mathrm{\Delta }{\hat{\tau }}^{i,\left(APA\right)}_{x,k}\left(\cdot \right)$ is the operator to obtain the absolute code phase error with the geometry distance prediction (i.e., the APA process) and the error models, and its analytical expression is defined as 

\begin{equation}\label{GrindEQ__15_}
\mathrm{\Delta }{\hat{\tau }}^{i,\left(APA\right)}_{r,k}\left({\hat{r}}^i_{x,k-1}\right)\triangleq \frac{f_c}{c}\left({\tilde{\rho }}^i_{r,k-1}-\left({\hat{r}}^i_{x,k-1}+\left({\hat{B}}_{r,\rho ,t,k-1}+{\hat{B}}^i_{\rho ,sys,k-1}\right)-{\kappa }_D{\hat{B}}_{r,mp,k-1}\right)\right)
\end{equation}

Therefore, based on these discussions, it is easy to find that the absolute code phase error estimate $\mathrm{\Delta }{\hat{\tau }}^{i+,\left(RTK/INS\right)}_{code,0,k}$ in \eqref{GrindEQ__4_} (i.e., the difference between the received initial code phase and the counterpart of the local code replica synthesized with the NCO) can be alleviated by the proposed algorithm. 

Finally, the proposed algorithm in this paper is summarized in \textbf{Algorithm 1}. This algorithm is realized in a GPS SDR prototype where L1 C/A signals are used to validate the TOA and position estimation performance. 

\begin{table*}[htbp]
	\begin{tabular}{ll}
	\toprule
	\multicolumn{2}{c}{\textbf{Algorithm 1 }High-accuracy APA GNSS code phase tracking based on RTK/INS deep integration} \\ 
	\midrule
	\textbf{Require:} & $k^*\triangleq k\ \mathrm{mod}\ KM$, subject to $K,M\in {\mathbb{Z}}^+$ and $k,k^*\in \mathbb{N}$ \\ 
	1: & \makecell[l]{\textbf{while }new digital IF samples (for a coherent processing interval) are received at the $k$th \\epoch \textbf{do}} \\  
	2: &  \:\:\:\:Synthesize the code and carrier local replicas with the code/carrier NCOs;  \\  
	3: &     \:\:\:\:\makecell[l]{Produce the early- prompt- and late-branch samples through the I\&D using the local replicas \\and the received IF samples;}  \\  
	4: &     \:\:\:\:Discriminate the code/carrier phase errors with the outputs of the I\&D (i.e., correlator outputs);  \\  
	5: &     \:\:\:\:\textbf{if} the base station information is available at the tracking epoch(s) $\left\{k^*-2M,\ \dots ,k^*-M-1\right\}$ \textbf{then} \\  
	6: &         \:\:\:\:\:\:\:\:Compensate for the discriminated code phase error with \eqref{GrindEQ__14_}; \\  
	7: &     \:\:\:\:\textbf{else if }the base station information is available at the tracking epoch(s) $\left\{k^*-M,\dots ,\ k^*-1\right\}$ \textbf{then} \\  
	8: &         \:\:\:\:\:\:\:\:Compensate for the discriminated code phase error with \eqref{GrindEQ__12_}; \\  
	9: &     \:\:\:\:\textbf{else } \\  
	10: &         \:\:\:\:\:\:\:\:Compensate for the discriminated code phase error with \eqref{GrindEQ__13_}; \\  
	11: &     \:\:\:\:\textbf{end if} \\  
	12: &     \:\:\:\:Optimize the compensated code phase error from \textbf{Step 6/8/10} with a 1-Hz 2nd-order loop filter; \\  
	13: &     \:\:\:\:\textbf{if} the vector tracking trigger (5 Hz) is activated \textbf{then} \\  
	14: &         \:\:\:\:\:\:\:\:Optimize the discriminated carrier phase error with a 0.5-Hz 1st-order loop filter; \textbf{} \\  
	15: &         \:\:\:\:\:\:\:\:\textbf{if} the RTK/INS EKF is updated at the epoch(s) $\left\{k^*,k^*-2M,k^*-3M,\dots ,k^*-\left(K-1\right)M\right\}$ \textbf{then} \\  
	16: &             \:\:\:\:\:\:\:\:\:\:\:\:Predict the carrier Doppler with \eqref{GrindEQ__10_}; \\  
	17: &         \:\:\:\:\:\:\:\:\textbf{else if} the RTK/INS EKF is updated at the epoch(s) $\left\{k^*-2M+1,\dots ,k^*-M-1\right\}$ \textbf{then} \\  
	18: &             \:\:\:\:\:\:\:\:\:\:\:\:Predict the carrier Doppler with \eqref{GrindEQ__8_}; \\  
	19: &         \:\:\:\:\:\:\:\:\textbf{else if} the RTK/INS EKF is updated at the epoch(s) $\left\{k^*-M\right\}$ \textbf{then} \\  
	20: &             \:\:\:\:\:\:\:\:\:\:\:\:Predict the carrier Doppler with \eqref{GrindEQ__9_}; \\  
	21: &         \:\:\:\:\:\:\:\:\textbf{else if }the RTK/INS EKF is updated at the epoch(s) $\left\{k^*-M+1,\dots ,k^*-1\right\}$ \textbf{then} \\  
	22: &             \:\:\:\:\:\:\:\:\:\:\:\:Predict the carrier Doppler with \eqref{GrindEQ__7_}; \\  
	23: &         \:\:\:\:\:\:\:\:\textbf{else} \\  
	24: &             \:\:\:\:\:\:\:\:\:\:\:\:Predict the carrier Doppler with \eqref{GrindEQ__11_}; \\  
	25: &         \:\:\:\:\:\:\:\:\textbf{end if} \\  
	26: &         \:\:\:\:\:\:\:\:Compute the carrier frequency with \eqref{GrindEQ__6_} (for carrier NCO); \\  
	27: &     \:\:\:\:\textbf{else } \\  
	28: &         \:\:\:\:\:\:\:\:Optimize and predict the carrier Doppler with a 15-Hz 3rd-order loop filter (for carrier NCO); \\  
	29: &     \:\:\:\:\textbf{end if} \\  
	30: &     \:\:\:\:Compute the code frequency with \eqref{GrindEQ__5_} (for code NCO);  \\  
	31: & \textbf{end while} \\  
	\bottomrule
	\end{tabular}
	\label{tab:alg1}
\end{table*}

\section{Results and Discussion}\label{sec3}

\begin{figure}[htbp]%
\centerline{\includegraphics[width=7.5cm]{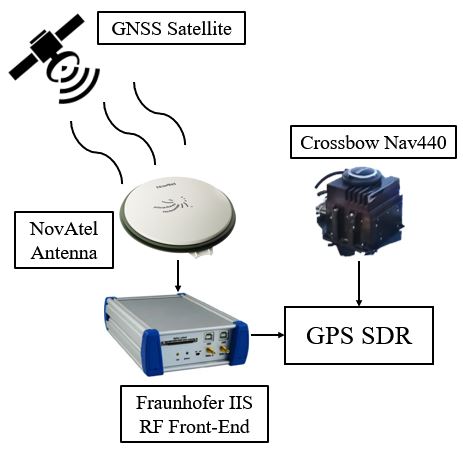}}
\caption{Setup for the stationary experiments. }
\label{fig:fig5}
\end{figure}

The experimental equipment is set up as shown in \reffig{fig:fig5}. Two stationary data sets were collected in the real world to verify the proposed algorithm. A NovAtel antenna was used to receive the GPS L1 C/A IF signals through a Fraunhofer IIS RF frond-end, where the IF sampling rate is 10.125 MHz. The IMU raw data were collected from the Crossbow Nav 440 device, where the IMU's gyro and accelerometer bias stabilities are 10 deg/h and 1 mg, respectively. Besides, it is worth mentioning that two samples are taken for updating the inertial sensor data for our navigation equation, so the updating rate of the INS DR is half (50 Hz) of the IMU raw data rate (100 Hz). The reference positions of the two experiments are obtained by averaging the results provided by the Crossbow Nav440 GPS/INS integration solutions (the centers of the IMU sensor and the GNSS antenna are sufficiently close in the setup and neglected in this experiment). 

The proposed algorithm is tested in a GPS SDR platform where the coherent integration time is 5 ms, the classic discriminators are chosen as the noncoherent-early-minus-late-amplitude code discriminator and Costas carrier discriminator, and the early-late spacing is four IF sample intervals. Five types of tracking algorithms are compared in the same SDR conditions except for the parameter adjustment in Table \ref{tab:tab1}. 

\begin{table}[htbp]%
	\centering
	\caption{Parameter settings of the SDR regarding the real-world experimental validation}
	\begin{tabular}{cccccc}
	\toprule
	\makecell{\textbf{SDR Type}} & \makecell{Relative \\position/ \\velocity \\aiding} & \makecell{RTK \\absolute \\position \\aiding} & \makecell{INS deep \\integration} & \makecell{RTK/INS \\integration \\navigation \\solution} & \makecell{Tracking \\loop filter} \\ 
	\midrule
	\makecell{STL (traditional) \\\cite{Kaplan2017}} & No & No & No & No & \makecell{1-Hz 2nd-order DLL \& \\18-Hz 3rd-order PLL} \\  
	\makecell{RTK-based VDFLL \\\cite{Lashley2021}} & Yes & No & No & No & \multirow{4}{*}{\makecell[c]{1-Hz 2nd-order DLL \& \\0.5-Hz 1st-order PLL \& \\15-Hz 3rd-order PLL \\(see \textbf{Algorithm 1})}} \\  
	\makecell{RTK/INS-based VDFLL \\\cite{Lashley2021}} & Yes & Yes & Yes & Yes &  \\  
	\makecell{RTK-based VDFPLL \\\cite{Luo2022aa}} & Yes & Yes & No & No &  \\  
	\makecell{RTK/INS-based VDFPLL \\(proposed)} & Yes & Yes & Yes & Yes &  \\  
	\bottomrule
	\end{tabular}
	\label{tab:tab1}
\end{table}

\begin{figure}[htbp]%
\centering
\subfloat{{\includegraphics[width=8cm]{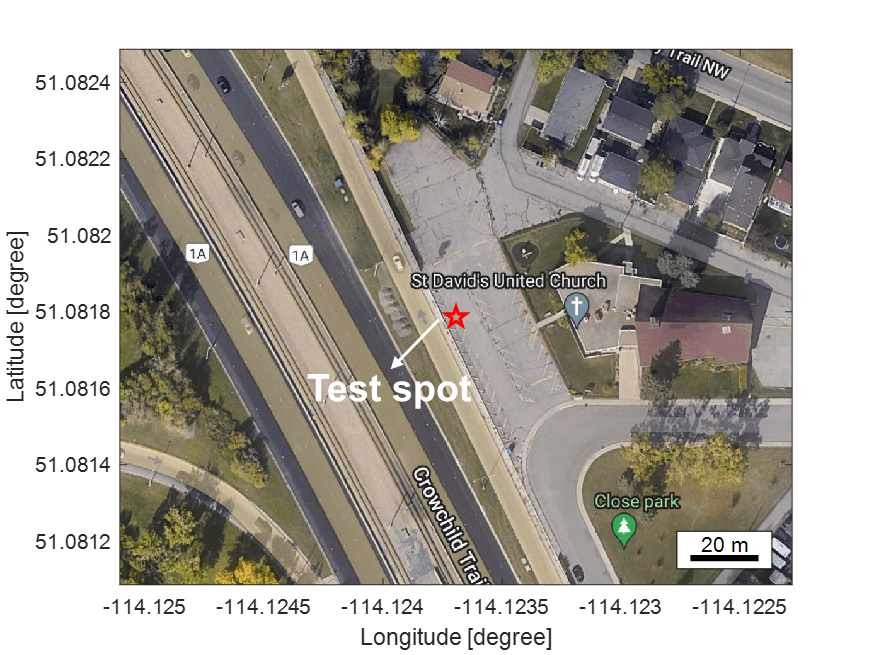}}}%
\hfil
\subfloat{{\includegraphics[width=7cm]{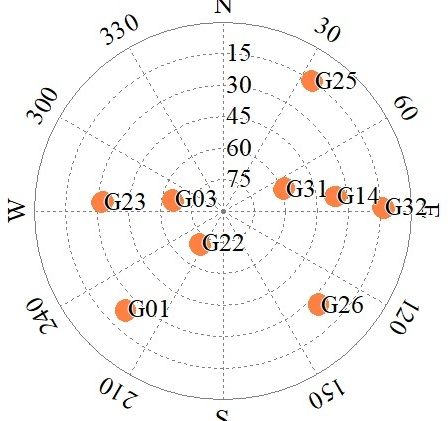}}}%
\caption{Open-sky test spot (Google Map show) and sky plot of available GPS satellites.}%
\label{fig:fig6}%
\end{figure}

First, an open sky area is chosen to carry out the experiment where the test spot in Google Map and the sky plot of the available GPS satellites are shown in \reffig{fig:fig6}.

\begin{figure}%
\centering
\subfloat {{\includegraphics[width=7.5cm]{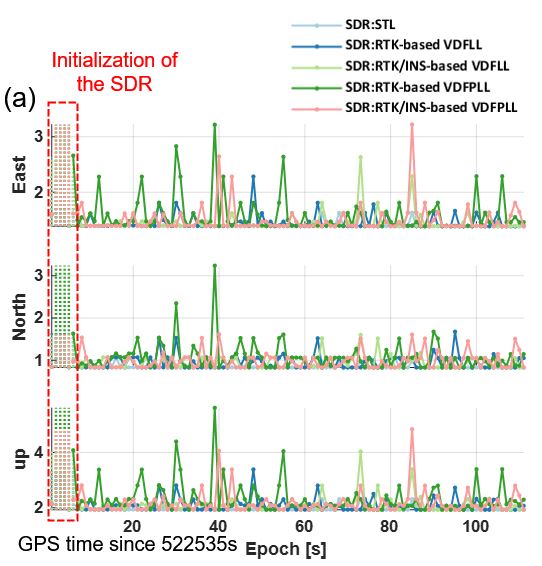}}}%
\hfil
\subfloat {{\includegraphics[width=8cm]{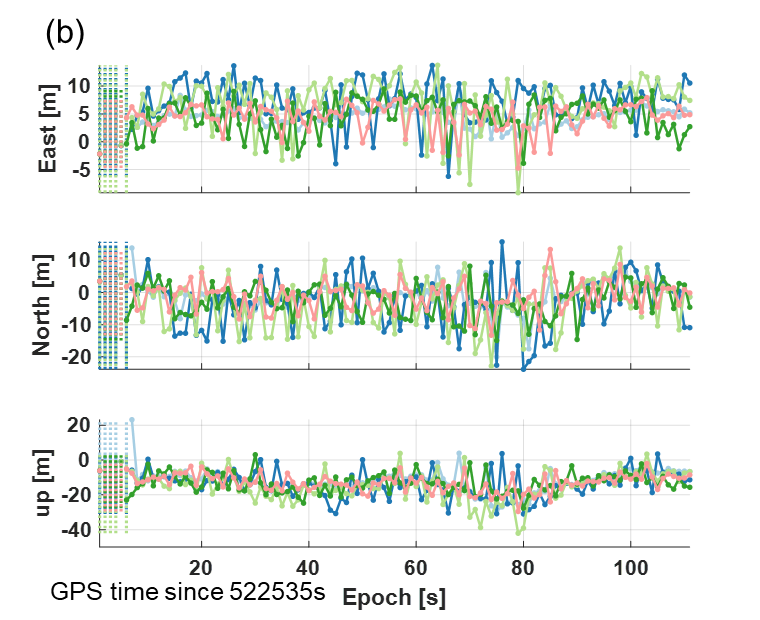}}}%
\hfil
\subfloat {\centerline{\includegraphics[width=6cm]{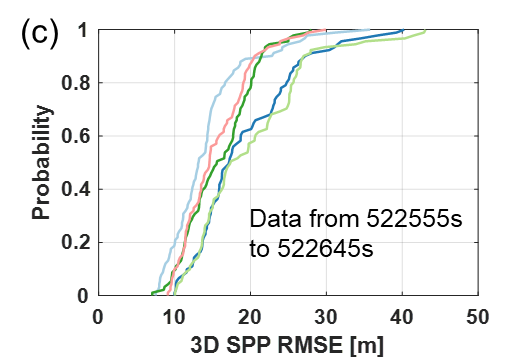}}}%
\caption{Single point navigation results and statistical analysis of different SDRs in the open-sky situation where dashed lines correspond to outlier epochs. (a) DOP values (b) SPP results (c) CDF curves of 3D SPP RMSE.}%
\label{fig:fig7}%
\end{figure}

We first assess the SPP results for this open-sky case. The SPP is based on the weighted NLS algorithm in the tested SDR, which refers to an open-source package RTKLIB \cite{Takasu2009}. \reffig{fig:fig7} depicts the dilution of precision (DOP) results, SPP errors, and the 3D position cumulative distribution function (CDF) curves of 3D SPP root-mean-squared errors (RMSEs). 

Although the results show that the proposed algorithm does not increase the SPP accuracy compared to the classic STL algorithm in the open sky, it proves that the APA-based VDFPLL algorithms enhance the traditional RPA-based VDFLL in the static situation. Meanwhile, using the INS can moderately enhance the RTK-based APA tracking process.

\begin{figure}[htbp]%
\centering
\subfloat {{\includegraphics[width=7.5cm]{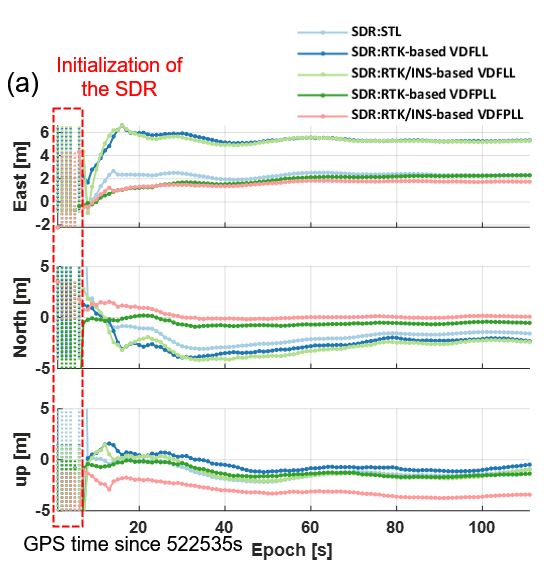}}}%
\hfil
\subfloat {{\includegraphics[width=7.5cm]{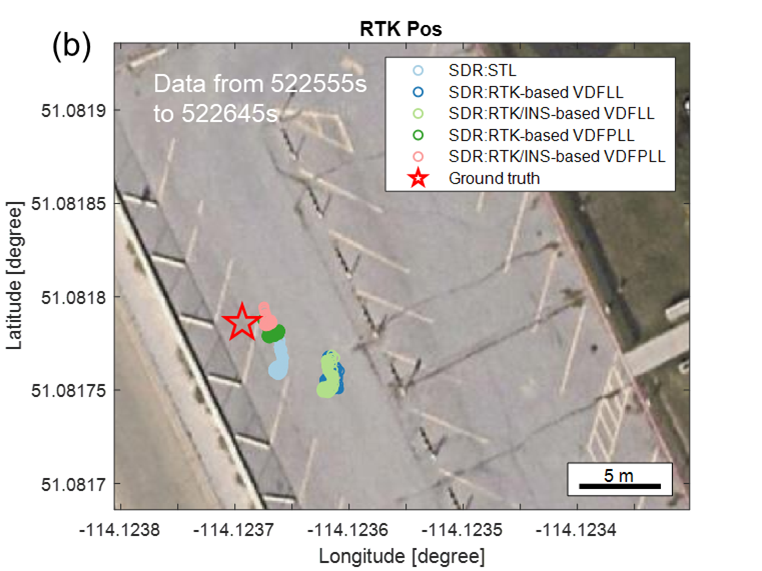}}}%
\hfil
\subfloat {{\includegraphics[width=6cm]{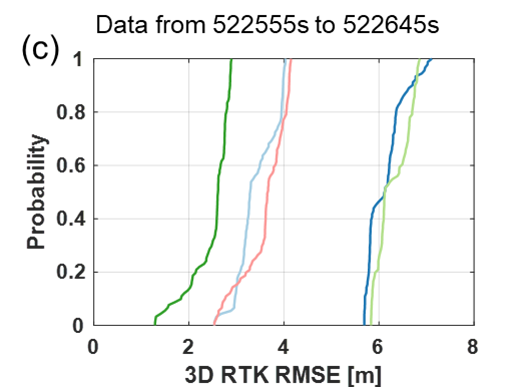}}}%
\hfil
\subfloat {{\includegraphics[width=6cm]{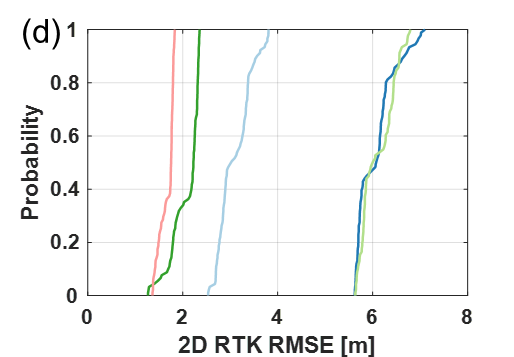}}}%
\caption{RTK position results and statistical analysis of different SDRs in the open-sky situation where dashed lines correspond to the outlier epochs. (a) RTK position errors (b) horizontal RTK results in Google Map (c) CDF curves of 3D RTK RMSE (d) CDF curves of horizontal (2D) RTK RMSE. }%
\label{fig:fig8}%
\end{figure}

Then, the RTK position errors for the different SDR algorithms are compared in \reffig{fig:fig8}, where the position errors, horizontal position results for a Google Map show, and the CDF curves of 3D and 2D position estimates RMSE are included. The traditional RPA VDFLL does not help the RTK position accuracy in this open-sky and static case. The finding is that the RTK-only-based VDFPLL improves the RTK results within a 3D range while the RTK/INS-based VDFPLL slightly makes it outperform the STL-based horizontal positioning. It can be explained that the vertical INS DR position solution would cause more code phase errors (compared to the STL tracking in a static open-sky case) in terms of the vertical position estimation. 

\begin{figure}[htbp]%
\centering
\subfloat {{\includegraphics[width=7.5cm]{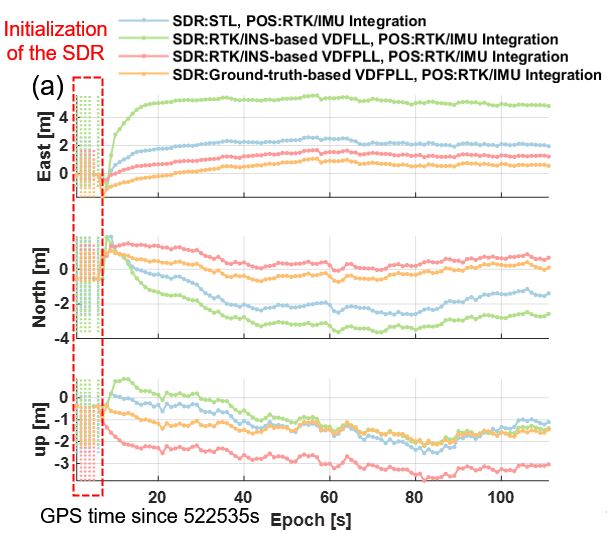}}}%
\hfil
\subfloat {{\includegraphics[width=7.5cm]{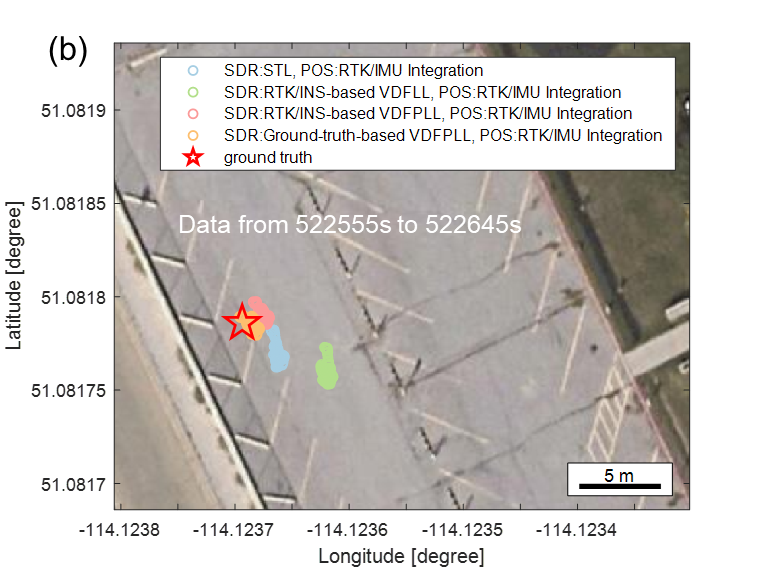}}}%
\hfil
\subfloat {{\includegraphics[width=6cm]{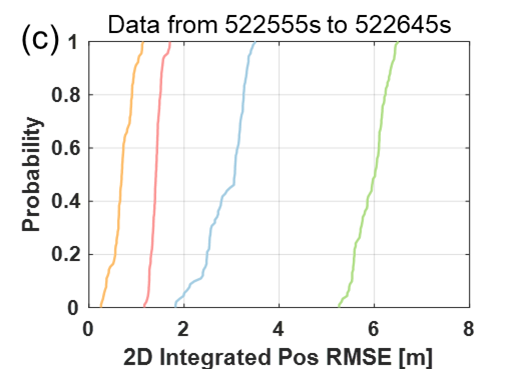}}}%
\hfil
\subfloat {{\includegraphics[width=6cm]{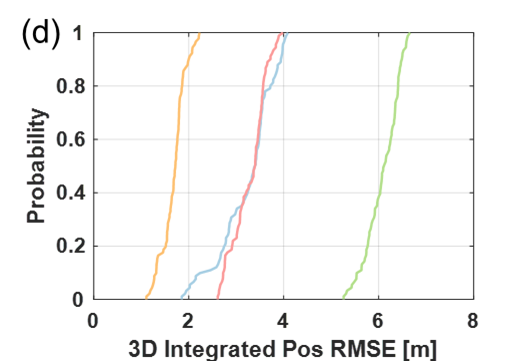}}}%
\caption{RTK position results and statistical analysis of different SDRs in the open-sky situation where dashed lines correspond to the outlier epochs. (a) RTK position errors (b) horizontal RTK results in Google Map (c) CDF curves of 3D RTK RMSE (d) CDF curves of horizontal (2D) RTK RMSE. }%
\label{fig:fig9}%
\end{figure}

Next, we compare the RTK/INS integrated results of three SDRs in \reffig{fig:fig9}, where ``Ground-truth-based VDFPLL'' means that the SDR leverages the actual position coordinates instead of on-the-fly RTK or integrated RTK/INS position estimates. It is a hardware-in-the-loop (HIL) simulation strategy to provide a reference for the proposed algorithm under the same SDR conditions. In other words, the HIL simulation results represent the upper bound of the performance that the used SDR platform can achieve. 

Based on the CDF curves in \reffig{fig:fig9}(d), the proposed accuracy slightly exceeds the traditional STL-based integrated RTK/INS solution within the 78\% probability. So, the large errors can be alleviated in the integration results using the proposed algorithm in the open sky area. In contrast, the traditional VDFLL-based integrated RTK/INS positioning accuracy drops much in this well-condition static scenario.
 
TOA curves will be evaluated in the following parts to examine the baseband processing discrepancy. The TOA estimation from the GPS SDR is computed via \eqref{GrindEQ__3_}, where it is worth noting that the ${\tilde{\rho }}^i_k$ here is the raw data from the loop filter output without the carrier-smoothing algorithm. Therefore, the measurement and reference TOA residuals in meters are computed as follows
\[{\left(\mathrm{TOA\ measurement\ residual}\right)}^i_k\buildrel \Delta \over ={\tilde{\rho }}^i_k-c\left({TOA}^i_0+\delta {\hat{t}}_{r,0}\right)-c\left(-f^i_{d,0}f^{-1}_r+\delta {\hat{\dot{t}}}_{r,0}\right)kT_{coh}\] 
\[{\left(\mathrm{TOA\ reference\ residual}\right)}^i_k\buildrel \Delta \over =c{TOA}^i_k-c{TOA}^i_0-c\left(-f^i_{d,0}f^{-1}_r\right)kT_{coh}\] 
\[\mathrm{TOA\ residual}\mathrm{\ error}\buildrel \Delta \over ={\left(\mathrm{TOA\ measurement\ residual}\right)}^i_k-{\left(\mathrm{TOA\ reference\ residual}\right)}^i_k\] 
where the TOA residuals exclude the initial pseudorange and initial Doppler frequency for the simplicity of analysis; subscript $0$ and $k$ denote the initial and the $k$th epoch, respectively; ${TOA}^i_0$ and $f^i_{d,0}$ represent the TOA and Doppler frequency reference at the initial epoch, and they are computed as
\[{TOA}^i_0\mathrm{=}c^{-1}\left(\left\|{\boldsymbol{\mathrm{p}}}_{gt,0}-{\boldsymbol{\mathrm{p}}}^i_0\right\|+{\hat{B}}_{I,0}+{\hat{B}}_{T,0}-c\delta {\hat{t}}^i_0\right)\] 
\[f^i_{d,0}=\frac{f_r}{c}\left({\boldsymbol{\mathrm{v}}}_{gt,0}\cdot {\hat{\boldsymbol{\mathrm{e}}}}^i_0-{\hat{\boldsymbol{\mathrm{v}}}}^i_0\cdot {\hat{\boldsymbol{\mathrm{e}}}}^i_0-c\delta {\hat{\dot{t}}}^i_0\right)\] 
where ${\boldsymbol{\mathrm{p}}}_{gt,0}$ and ${\boldsymbol{\mathrm{v}}}_{gt,0}$ are the ground truth vectors in the ECEF coordinate frame for the user's position and velocity in the stationary experiments, with ${\boldsymbol{\mathrm{v}}}_{gt,0}={\left[0,0,0\right]}^T$; ${\boldsymbol{\mathrm{p}}}^i_0$, ${\boldsymbol{\mathrm{v}}}^i_0$, $\delta {\hat{t}}^i_0$, and $\delta{\hat{\dot{t}}}^i_0$ are the satellite position and velocity vectors, satellite clock bias and drift errors, respectively, which are obtained and computed using the broadcast ephemeris; meanwhile, ${\hat{B}}_{I,0}$ and ${\hat{B}}_{T,0}$ are the ionospheric error derived from the Klobuchar model and the tropospheric error calculated via the Saastamoninen model; ${\hat{\boldsymbol{\mathrm{e}}}}^i_0$ is the unit cosine vector computed from ${\boldsymbol{\mathrm{p}}}_{gt,0}$; $\delta {\hat{t}}^i_0$ and $\delta {\hat{\dot{t}}}^i_0$ are the estimated user's clock bias and drift errors computed as
\[\delta {\hat{t}}_{r,0}=averaging\ clock\ bias\ error-\frac{\left(navigation\ time\ spanning\right)}{2}\times \delta {\hat{\dot{t}}}_{r,0}\] 
where $\delta {\hat{\dot{t}}}_{r,0}$ is the averaging value of the clock drift estimates from the ``Ground-truth-based VDFPLL'' SDR. 

It is worthwhile to say that the given $\delta {\hat{\dot{t}}}^i_0$ and $\delta {\hat{t}}^i_0$ references are not sufficiently accurate, but they are satisfactory in validating the TOA performance amid different SDR algorithms. 

\begin{figure}[htbp]%
\centering
\subfloat {{\includegraphics[width=7cm]{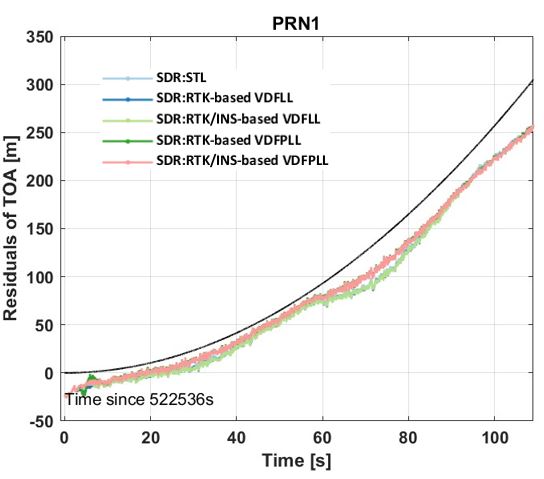}}}%
\hfil
\subfloat {{\includegraphics[width=7.5cm]{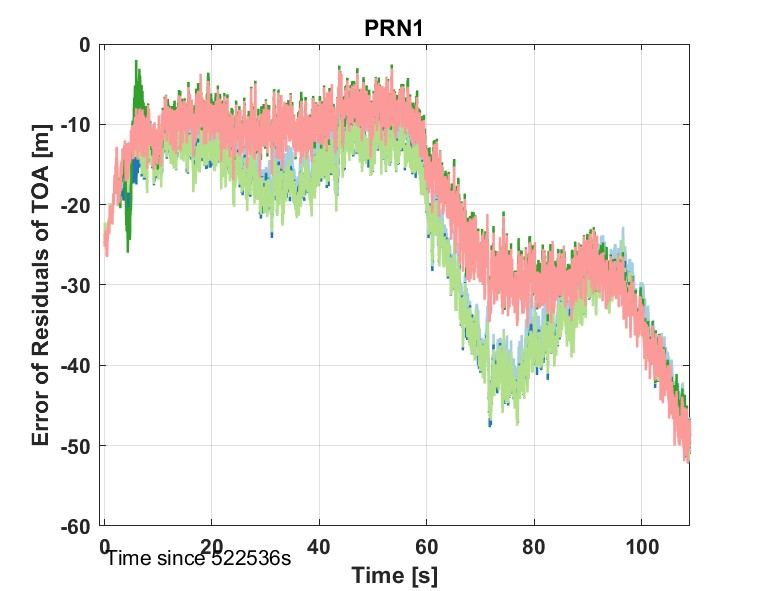}}}%
\hfil
\subfloat {{\includegraphics[width=7.5cm]{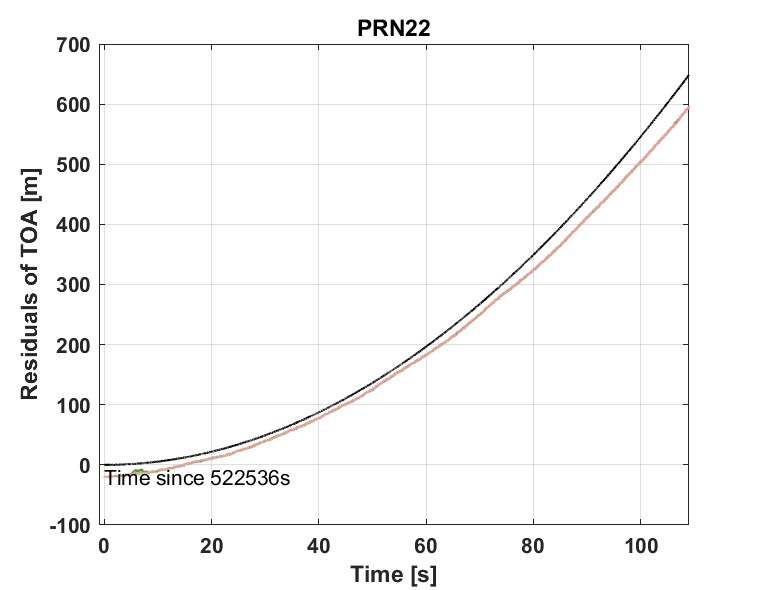}}}%
\hfil
\subfloat {{\includegraphics[width=7.5cm]{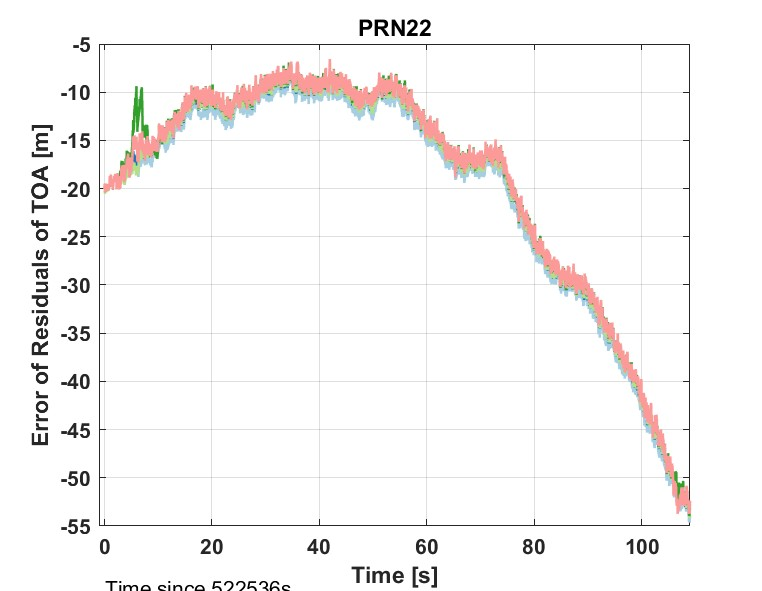}}}%
\caption{Error curves of the TOA residuals for the GPS satellites PRN1 and PRN22 in the open-sky situation. }%
\label{fig:fig10}%
\end{figure}

\reffig{fig:fig10} shows the error curves of the TOA residuals for the signals from the satellite of PRN1 (low elevation angle) and PRN22 (high elevation angle). It proves that the APA-based vector tracking algorithms perform better in interference mitigation for the signals from the low satellite. 

In summary, even if the proposed algorithm does not manifest much better than the traditional STL in an open-sky static environment, it demonstrates a significant improvement in comparison with the traditional RPA vector tracking in the same condition. More specifically, it can be used as boosting complement for the existing vector GNSS receivers. 

\begin{figure}[htbp]%
\centering
\subfloat {{\includegraphics[width=8cm]{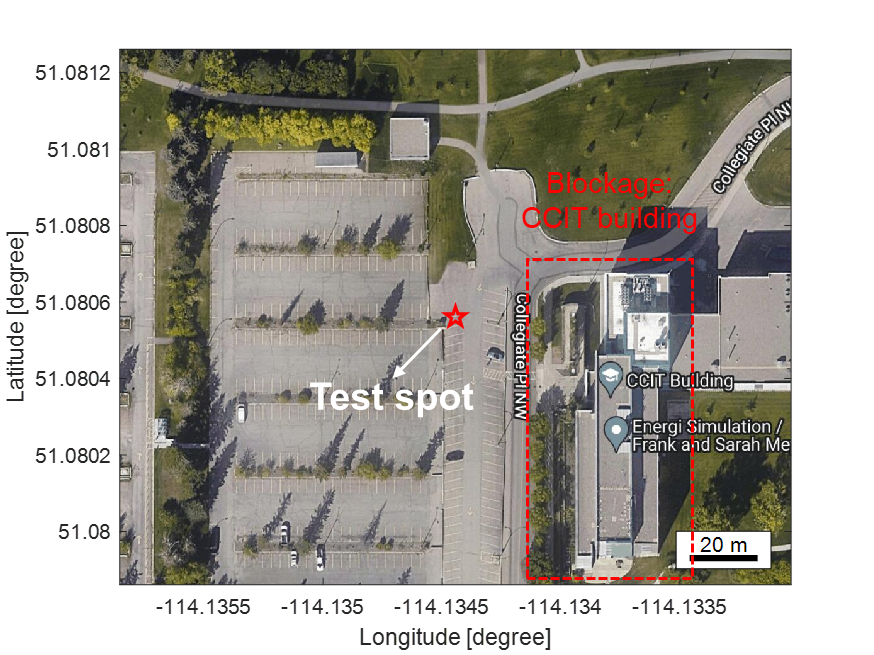}}}%
\hfil
\subfloat {{\includegraphics[width=7cm]{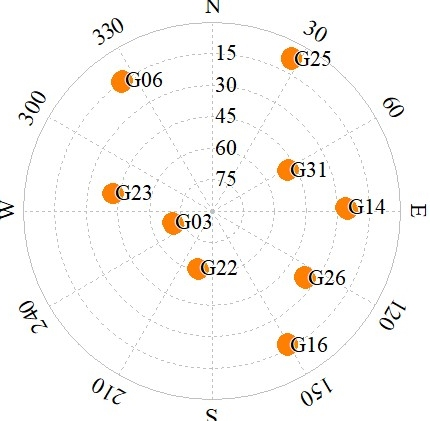}}}%
\caption{Semi-open-sky test spot (Google Map show) and the sky plot of available GPS satellites. }%
\label{fig:fig11}%
\end{figure}

Another set of data was collected under a semi-open-sky situation where the GPS antenna was receiving the signals affected by the eastern CCIT building at the campus of the University of Calgary. The Google Map show of the test spot and the corresponding satellite sky plot are provided in \reffig{fig:fig11}. 

\begin{figure}[htbp]%
\centering
\subfloat {{\includegraphics[width=7.5cm]{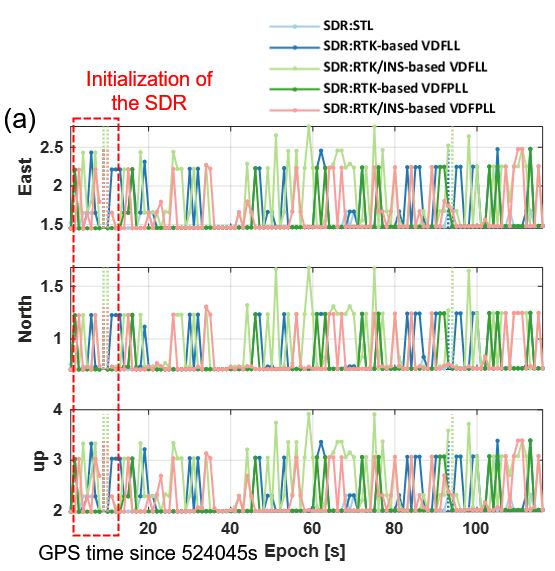}}}%
\hfil
\subfloat {{\includegraphics[width=8cm]{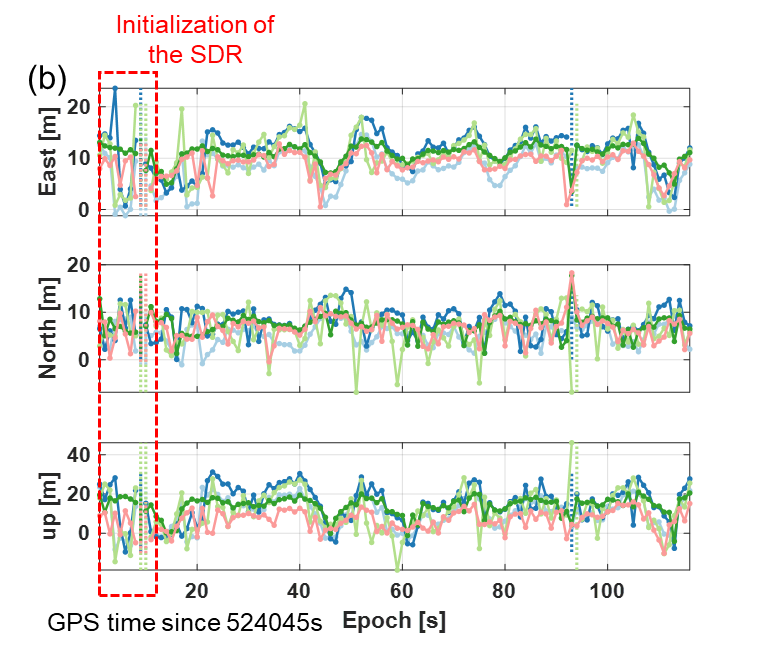}}}%
\hfil
\subfloat {\centerline{\includegraphics[width=6cm]{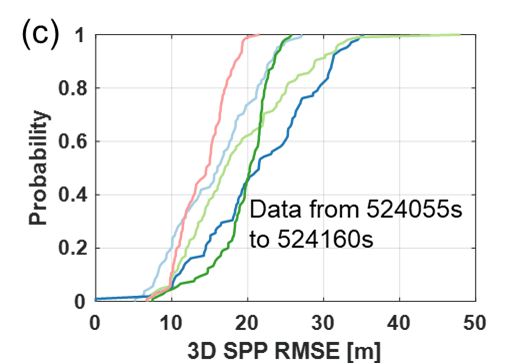}}}%
\caption{Single point navigation results and statistical analysis of different SDRs in the semi-open-sky situation where dashed lines correspond to the outlier epochs. (a) DOP values (b) SPP results (c) CDF curves of 3D SPP RMSE. }%
\label{fig:fig12}%
\end{figure}

In this test, where the results are displayed in \reffig{fig:fig12}, we also give the DOP values to offer the satellite geometry status; the SPP error curves and their 3D CDF are provided as well. 

First, it can be observed that the two VDFPLLs and the STL produce more reliable solutions as the position outliers occur to the two VDFLLs at around the 95th epoch. Then, it is evident that the proposed algorithm embraces the highest SPP accuracy. 

\begin{figure}[htbp]%
\centering
\subfloat {{\includegraphics[width=7.5cm]{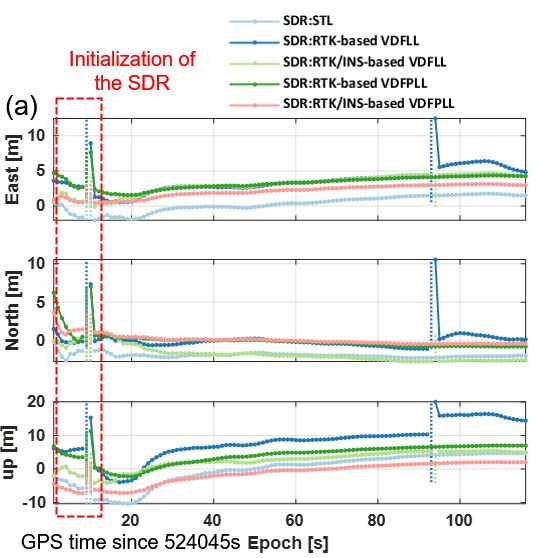}}}%
\hfil
\subfloat {{\includegraphics[width=8cm]{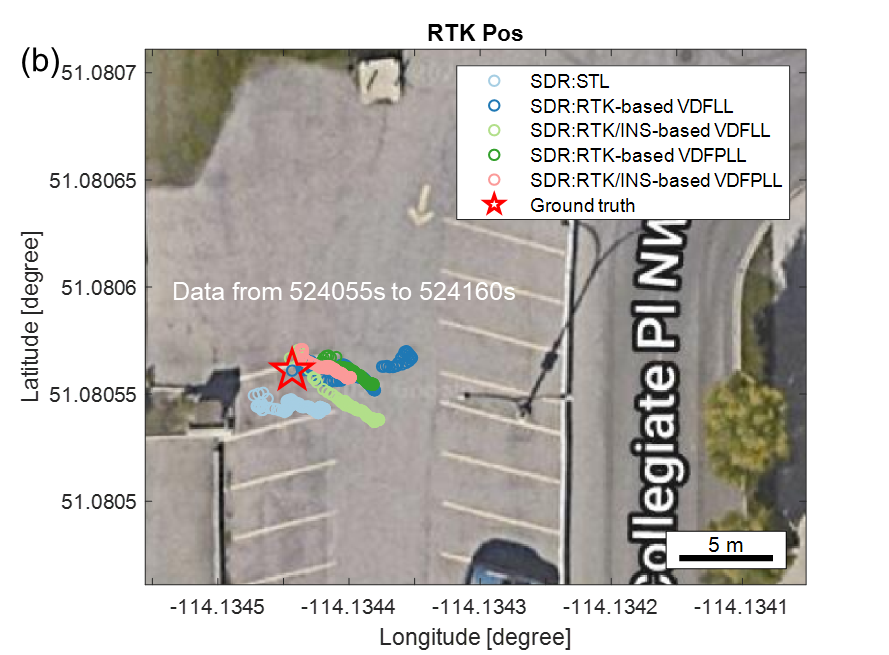}}}%
\hfil
\subfloat {\centerline{\includegraphics[width=6cm]{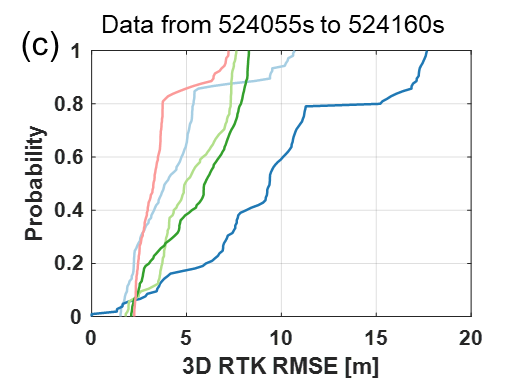}}}%
\caption{RTK position results and statistical analysis of different SDRs in the semi-open-sky situation where dashed lines correspond to the outlier epochs. (a) RTK position error (b) horizontal RTK position results in Google Map (c) CDF curves of 3D RTK RMSE. }%
\label{fig:fig13}%
\end{figure}

We also assess the RTK solution accuracy related to the different SDRs. The RTK position error curves, position results in Google Map, and the 3D CDF curves are provided in \reffig{fig:fig13}. After the SDR RTK solutions become stable, the RTK accuracy from the proposed SDR solutions still shows the highest performance. The RTK-only-based APA algorithm can reduce the random noise, but it is more biased than the traditional STL algorithm. The two RPA-only vector tracking techniques are still less capable of offering efficient assistance in the static test.  

\begin{figure}[htbp]%
	\centering
	\subfloat{{\includegraphics[width=7.5cm]{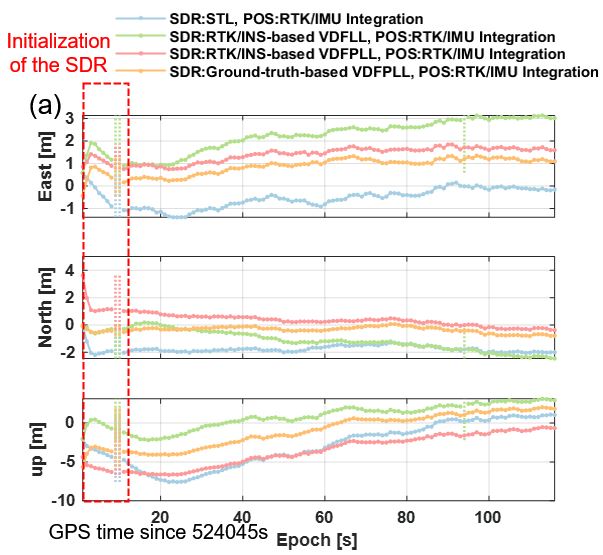}}}%
	\hfil
	\subfloat{{\includegraphics[width=8cm]{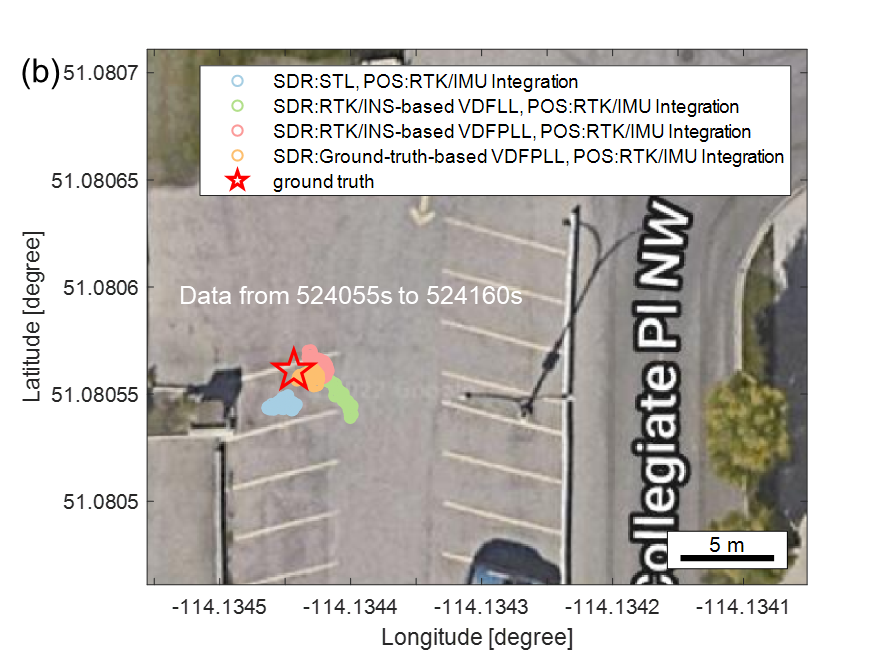}}}%
	\hfil
	\subfloat{{\includegraphics[width=6cm]{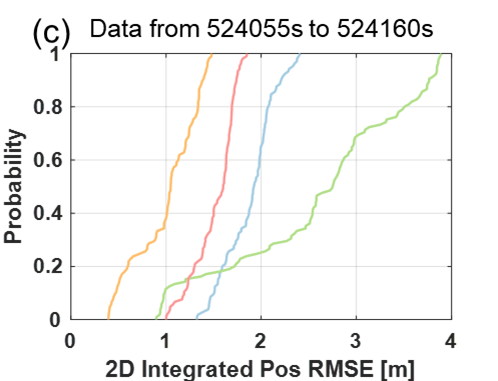}}}%
	\hfil
	\subfloat{{\includegraphics[width=6cm]{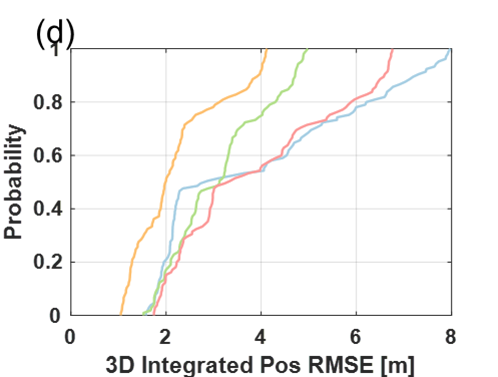}}}%
	\caption{RTK/INS integration position results and statistical analysis of different SDRs where dashed lines correspond to the outlier epochs in the semi-open-sky situation. (a) RTK/INS integration position error (b) horizontal RTK/INS integration position results in Google Map (c) CDF curves of 2D RTK/INS integration position RMSE (d) CDF curves of 3D RTK/INS integration position RMSE. }%
	\label{fig:fig14}%
\end{figure}

Next, the integrated RTK/INS solutions are compared in the semi-open-sky environment as shown in \reffig{fig:fig14}. In this case, the proposed algorithm significantly improved the 2D positioning performance. At the same time, the RTK/INS-based RPA method elevates the 3D positioning accuracy compared to the STL-based integration. The RPA- and APA-based integrated navigation accuracies over the error range of approximately 55\% probability outperform the traditional STL one. Furthermore, the positioning results with the proposed algorithm perform more stable than the traditional vector tracking. Another finding is that RTK/INS-based APA vector tracking yields a much more ideal horizontal positioning estimate than the RPA one. By contrast, the latter is superior to the former in the vertical direction. By comparing to the upper bound (i.e., the estimation from the ground-truth-based VDFPLL SDR), it can be explained that the fusion of the low-cost IMU has a side effect on the vertical position solution when it is applied to the proposed RTK/INS-based VDFPLL SDR in this stationary experiment.

\begin{figure}[htbp]%
\centering
\subfloat {{\includegraphics[width=7cm]{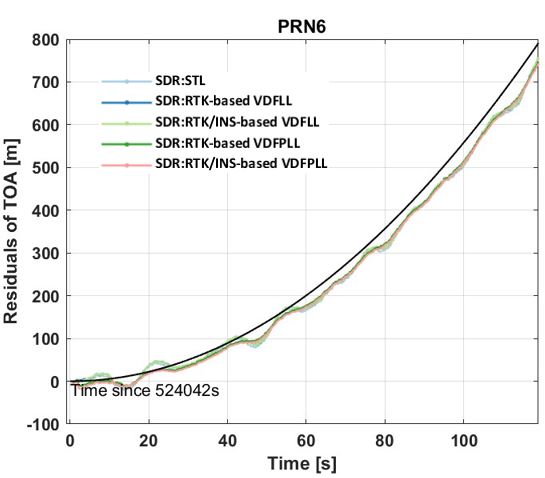}}}%
\hfil
\subfloat {{\includegraphics[width=7.5cm]{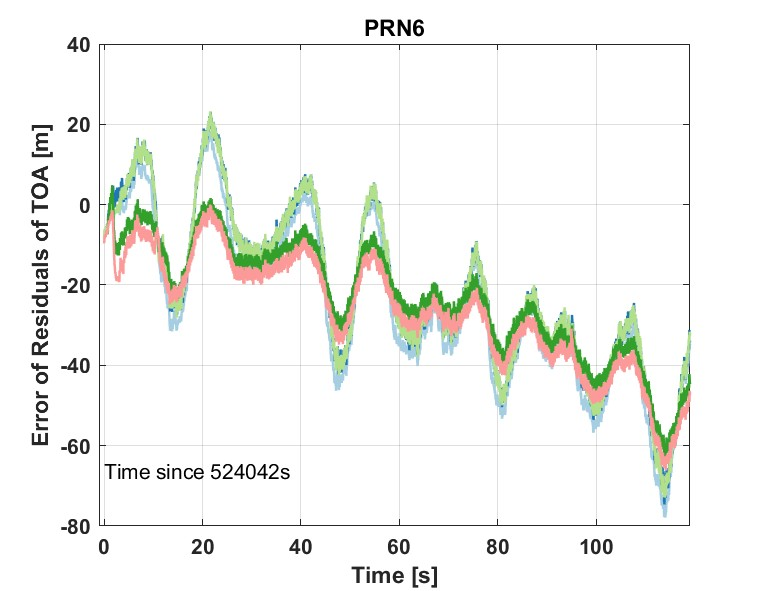}}}%
\hfil
\subfloat {{\includegraphics[width=7.5cm]{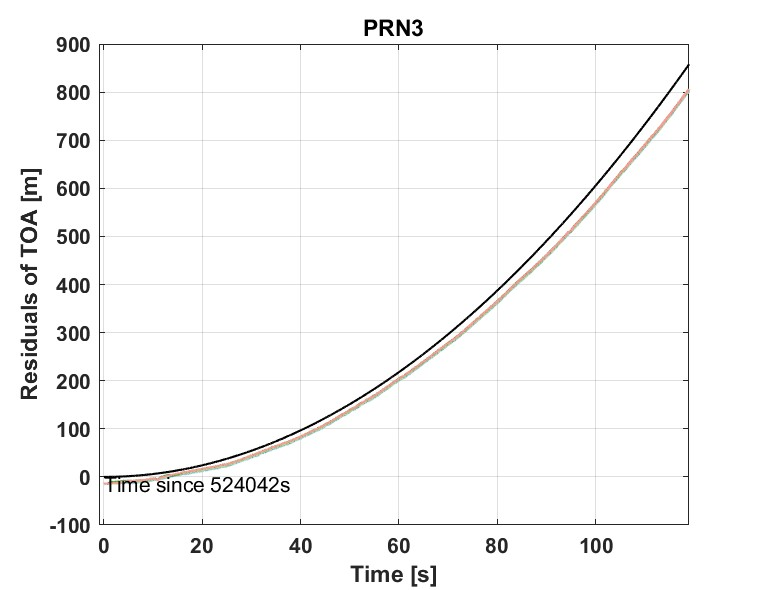}}}%
\hfil
\subfloat {{\includegraphics[width=7.5cm]{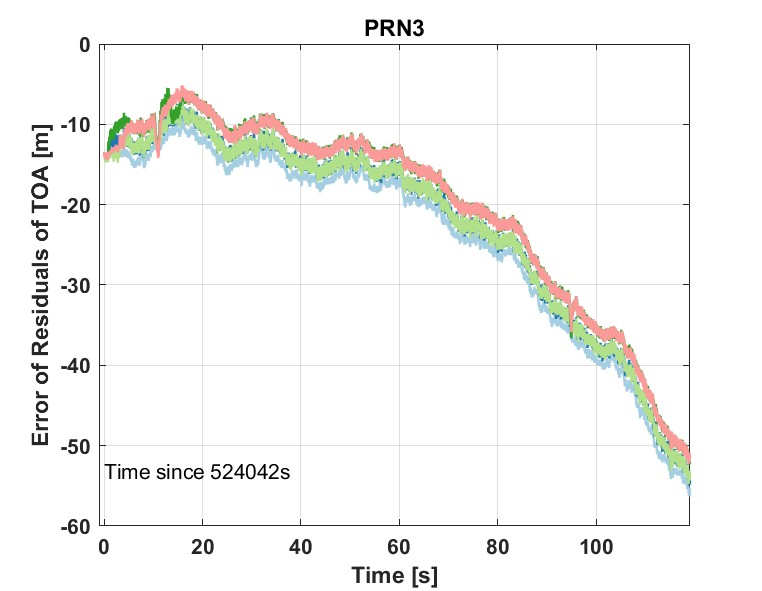}}}%
\caption{Error curves of the TOA residuals for the GPS satellites PRN6 and PRN3 in the semi-open-sky situation. }%
\label{fig:fig15}%
\end{figure}

Then, the raw TOA performance in terms of the signal from the high-elevation-angle satellite (PRN3) and the one with a low elevation angle affected by the multipath interference (PRN6) are plotted in \reffig{fig:fig15}. A more significant fluctuation in the TOA curves emerges at the PRN6 in this semi-open-sky experiment compared to the open-sky PRN1 (see \reffig{fig:fig10}). Also, both APA-based tracking loops are more capable of alleviating the TOA error varying with a long-term time spanning (i.e., the level of dozens of seconds) than the RPA-based vector tracking and the STL. Nevertheless, the curves computed from the high-quality signal, PRN3, are highly homogeneous regarding all the tested tracking loop algorithms.

The next part will quantitatively examine the exact improvement the proposed algorithm can offer for the GNSS baseband estimation. As mentioned, an upper bound of the instantaneous integrated positioning performance is obtained from the SDR under the HIL test using the ``Ground-truth-based VDFPLL''. Therefore, the corresponding TOA error curve representing the upper bound can also be extracted from the tracking results. Then, we compute the TOA error of the PRN22 and PRN3 (with the highest elevation angles during the experiments) in the open-sky and semi-open-sky cases as the respective references. The proposed TOA error references reasonably model the remained local clock errors in meters varying with the time where the other biased errors, like the atmospheric delay and initial TOA errors, are assumed to be well removed by the given models. 

\begin{figure}[htbp]%
\centering
\subfloat {{\includegraphics[width=7.5cm]{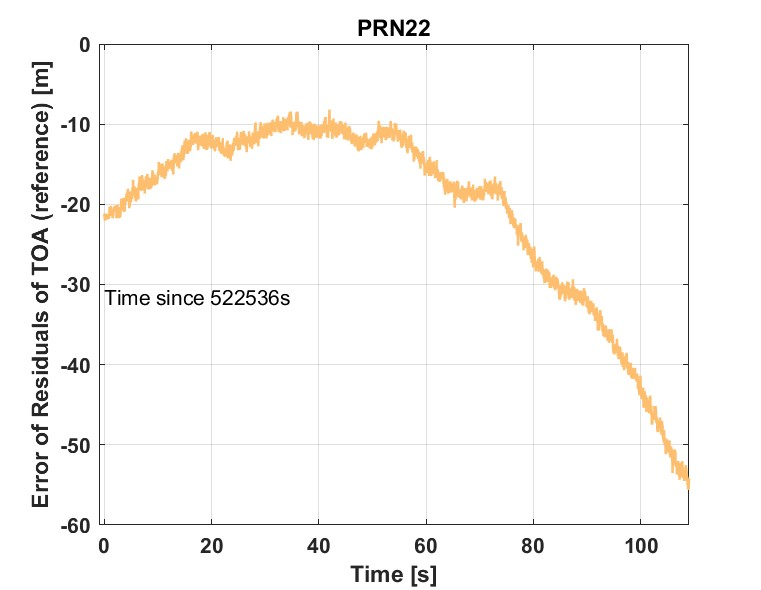}}}%
\hfil
\subfloat {{\includegraphics[width=7.5cm]{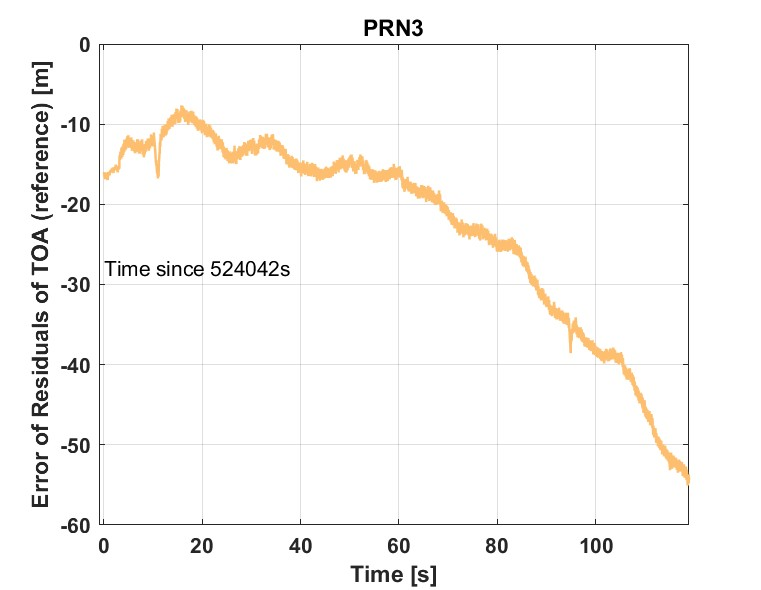}}}%
\caption{TOA curve references (regarding the error curves of the TOA residuals) derived from the ground-truth-based VDFPLL SDR. (a) TOA reference from PRN22 for the open-sky experiment (b) TOA reference from PRN3 for the semi-open-sky experiment. }%
\label{fig:fig16}%
\end{figure}

Ultimately, the TOA curve references are derived and illustrated in \reffig{fig:fig16}. After that, the TOA accuracy of different satellites in the two testing situations will be analyzed via these references in the following. 

\begin{figure}[htbp]%
\centering
\subfloat {{\includegraphics[width=8cm]{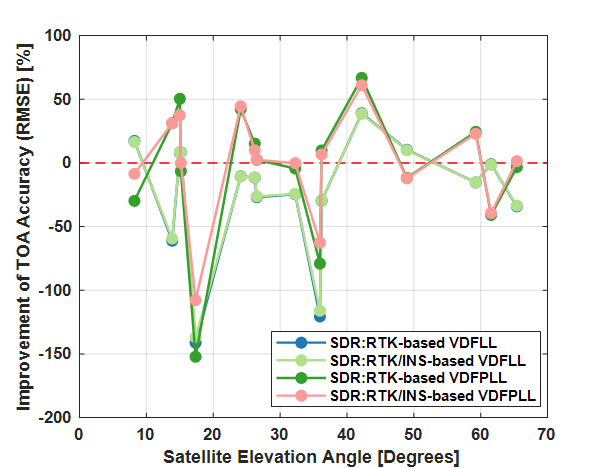}}}
\hfil
\subfloat {{\includegraphics[width=8cm]{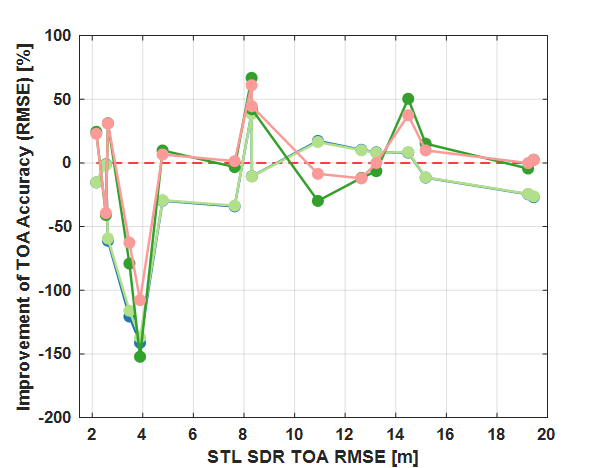}}}
\caption{Comparison of TOA accuracy improvements varying with the satellite elevation angles and the TOA errors (positive and negative values represent the improved and reduced performance percentages, respectively).}
\label{fig:fig17}
\end{figure}

Table \ref{tab:tab2} summarizes the statistical analysis of the TOA performances of different tracking algorithms where the RMSE results are computed for the used satellites. Then, in regard to the traditional STL, the TOA accuracy improvements of the two RPA- and APA-based vector tracking algorithms operated in the GPS SDR are computed and depicted in \reffig{fig:fig17}. 

\begin{table}[htbp]%
	\centering
	\caption{RMSEs of the TOA estimates from the active satellites regarding the two stationary experiments where the reference curves for the error of residuals of TOA refer to \reffig{fig:fig16} (``OS'' and'' SOS'' correspond to ``open-sky'' and ``semi-open-sky'' testing situations, respectively; ``Averaging $\mathrm{C/}{\mathrm{N}}_0$'' is estimated from the ``Ground-truth-based VDFPLL'' SDR).}
	\begin{tabular}{ccccccccc}

	\toprule
	\makecell{PRN \\numbers} & \makecell{Elevation \\angle [$\mathrm{{}^\circ}$]} & \makecell{Averaging \\$\mathrm{C/}{\mathrm{N}}_0$ \\\:[dB-Hz]} & \makecell{PRN \\numbers \\of the \\TOA error \\reference} & \multicolumn{5}{c}{\makecell[c]{RMSE for the error of residuals of TOA [m]}} \\

	\cmidrule{5-9}
	 &  &  &  & STL & \makecell{RTK \\VDFLL} & \makecell{RTK/INS \\VDFLL} & \makecell{RTK \\VDFPLL} & \makecell{Proposed \\RTK/INS \\VDFPLL} \\ 
	\midrule

	SOS-25 & 8.2 & 43.4 & SOS-03 & 10.92 & 9.03 & 9.12 & 14.17 & 11.85 \\  
	OS-32 & 13.9 & 45.2 & OS-22 & 2.62 & 4.21 & 4.17 & 1.80 & 1.80 \\  
	SOS-06 & 15.0 & 39.6 & SOS-03 & 14.49 & 13.31 & 13.24 & 7.18 & 9.05 \\  
	OS-25 & 15.2 & 44.8 & OS-22 & 13.23 & 12.12 & 12.16 & 14.08 & 13.22 \\  
	SOS-16 & 17.4 & 43.5 & SOS-03 & 3.89 & 9.39 & 9.22 & 9.81 & 8.08 \\  
	OS-01 & 24.1 & 37.4 & OS-22 & 8.31 & 9.18 & 9.19 & 4.80 & 4.61 \\  
	SOS-14 & 26.2 & 44.2 & SOS-03 & 15.18 & 16.92 & 16.89 & 12.87 & 13.67 \\  
	OS-26 & 26.5 & 41.0 & OS-22 & 19.45 & 24.68 & 24.54 & 18.96 & 18.97 \\  
	OS-23 & 32.3 & 47.4 & OS-22 & 19.23 & 23.96 & 23.91 & 20.07 & 19.24 \\  
	SOS-26 & 36.0 & 46.8 & SOS-03 & 3.47 & 7.66 & 7.51 & 6.22 & 5.65 \\  
	OS-14 & 36.2 & 46.9 & OS-22 & 4.78 & 6.19 & 6.17 & 4.31 & 4.46 \\  
	SOS-23 & 42.2 & 45.6 & SOS-03 & 8.30 & 5.06 & 5.10 & 2.75 & 3.23 \\  
	SOS-31 & 49.0 & 51.4 & SOS-03 & 12.64 & 11.34 & 11.38 & 14.12 & 14.17 \\  
	OS-31 & 59.3 & 52.9 & OS-22 & 2.16 & 2.49 & 2.49 & 1.63 & 1.67 \\  
	SOS-22 & 61.6 & 48.2 & SOS-03 & 2.54 & 2.57 & 2.58 & 3.58 & 3.54 \\  
	OS-03 & 65.4 & 49.8 & OS-22 & 7.63 & 10.23 & 10.18 & 7.87 & 7.52 \\  
	\bottomrule

	\end{tabular}
	\label{tab:tab2}

\end{table}

The curves in \reffig{fig:fig17} indicate that both APA tracking methods outperform two RPA ones in elevating the TOA accuracy. Furthermore, regarding the lower-elevation satellites and the smaller-TOA-error channels, the TOA errors induced by the navigation results through the vector feedback procedure are more likely to drop in the proposed RTK/INS-based APA approach than in the RTK-only APA one.

\begin{figure}[htbp]%
\centering
\subfloat {{\includegraphics[width=7.8cm]{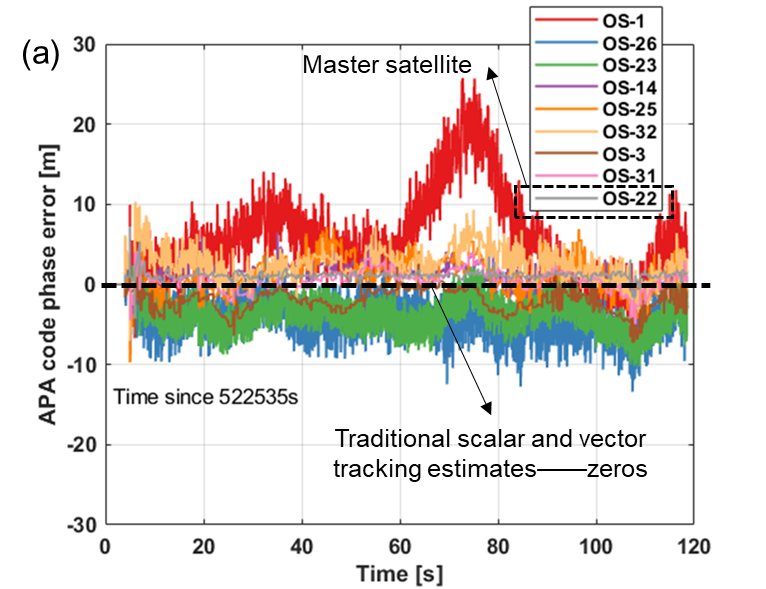}}}%
\hfil
\subfloat {{\includegraphics[width=7.8cm]{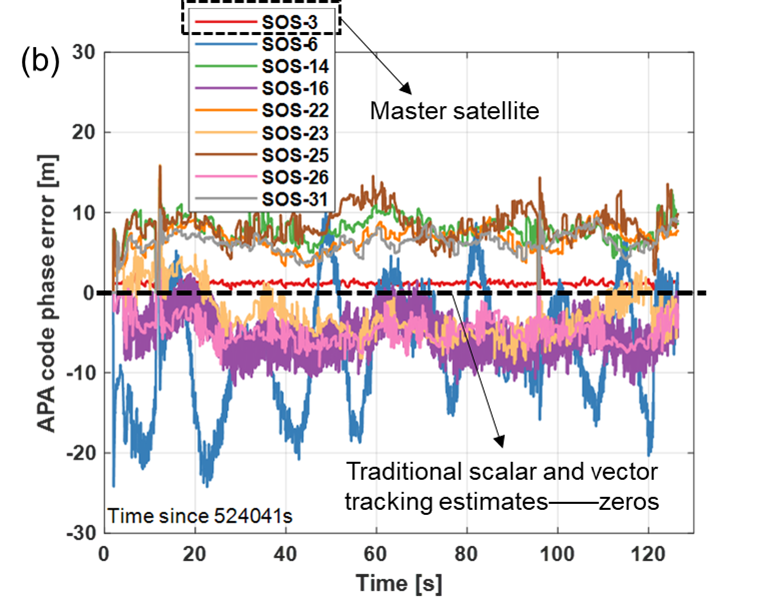}}}%
\caption{APA code phase errors from the proposed RTK/INS-based VDFPLL SDR where the numbers correspond to the satellite PRN numbers and dashed black lines correspond to the estimates from the traditional scalar and vector tracking algorithms (a) open sky (b) semi-open sky. }
\label{fig:fig18}
\end{figure}

\reffig{fig:fig18} plots the APA error curves estimated from the proposed RTK/INS-based VDFPLL modeling the instantaneous initial/absolute code phase error in meters at each tracking epoch of which the analytical expression is given by \eqref{GrindEQ__15_}. Compared to the traditional tracking algorithms (scalar and old vector tracking loops), the proposed algorithm can individually discriminate the absolute code phase error unrelating to the frequency error given by the same-epoch local replica subtracting incoming signals. This operation established through the proposed architecture is reasonable and it proves efficient. The implied information that the RTK/INS integrated EKF navigator provides more accurate positioning than the code-based-only SPP method can adequately explain the results. 

So, the dashed black lines in \reffig{fig:fig18} mean that the traditional scalar and vector tracking loops have nothing to recognize the code phase error not varying with the time spanning (the error residual remained by the traditional code discriminating process). However, the proposed RTK/INS-based APA vector tracking can directly estimate the absolute code phase error at every tracking epoch. The APA code discriminated results can show how the code phases are corrected by the accurate user's position solution, especially for the satellites facing the deterministic biased error changing as the cycles of dozens of seconds or longer. This phenomenon commonly occurs to the static user's antenna receiving incoming signals affected by the multipath effect. 

Finally, it is worth mentioning that the proposed RTK/INS-based VDFPLL is a simplified prototype applying the APA discriminated error relying on the INS and RTK to the GNSS baseband processing. The tracking performance still has space to be further improved by redoing loop filter algorithms (e.g., the EKF) or other GNSS baseband optimizing methods (e.g., snapshot processing \cite{Luo2022up,Fernandez-Hernandez2022} and open-loop tracking \cite{Tsang2022,VanGraas2009}). In other words, the proposed algorithm has a broad scope of use towards the GNSS signals at all frequencies and constellations, potentially contributing to the development of next-generation GNSS receivers and GNSS-based multi-sensor integrating navigation systems.

\section{Conclusions}\label{sec4}
This work proposes a deep integration of RTK and INS, enhancing the instantaneous code phase tracking performance in challenging static environments. In the presented algorithm, the navigation solutions, especially the absolute position solution, from the integrated EKF navigator are deeply fused into the GNSS tracking loop, forming an APA code phase discriminator. The RTK/INS-based APA discriminator combined with the vector tracking technique realized upon a GPS L1 C/A SDR can serve for more satisfactory tracking and positioning results than the RTK-based-only APA vector tracking approach. Two real-world stationary experiments have verified the performance. Finally, the conclusions of this work can be drawn as follows: 

\begin{enumerate}
\item  The proposed RTK/INS APA vector tracking has improved the multipath mitigation performance of the GNSS baseband in static situations compared to the traditional scalar/vector tracking and the RTK-aided-only APA vector tracking; 

\item  The deeply integrated INS in the proposed high-accuracy APA GPS SDR has enhanced the TOA estimation accuracy more significantly regarding the satellites with low elevation angles; 

\item  The technique regarding the tested low-cost IMU deeply integrated into the RTK-position-aided vector GPS has proved to be inferior in improving the vertical positioning accuracy but can efficiently increase the horizontal positioning accuracy in challenging static environments. 
\end{enumerate}

Our future work will focus on base-station-free APA GNSS tracking upon INS DR and precise point positioning (PPP) technique.

\bibliographystyle{unsrtnat}
\bibliography{references}  

\begin{thebibliography}{42}
\providecommand{\natexlab}[1]{#1}
\providecommand{\url}[1]{\texttt{#1}}
\expandafter\ifx\csname urlstyle\endcsname\relax
  \providecommand{\doi}[1]{doi: #1}\else
  \providecommand{\doi}{doi: \begingroup \urlstyle{rm}\Url}\fi

\bibitem[Sharma et~al.(2020)Sharma, Lichtenberger, and Pany]{Sharma2020}
H.~Sharma, C.~A. Lichtenberger, and T.~Pany.
\newblock {Multipath Error Modelling and Position Error Over-bounding for
  Precise RTK Positioning using GNSS Raw Measurements from Smartphone for
  Automotive Navigation}.
\newblock In \emph{Proceedings of the 33rd International Technical Meeting of
  the Satellite Division of The Institute of Navigation (ION GNSS+ 2020)},
  pages 1902--1924, oct 2020.
\newblock ISBN 0936406267.
\newblock \doi{10.33012/2020.17627}.
\newblock URL
  \url{https://www.ion.org/publications/abstract.cfm?articleID=17627}.

\bibitem[Zhang et~al.(2020)Zhang, Wen, Xu, and Hsu]{Zhang2020}
Guohao Zhang, Weisong Wen, Bing Xu, and Li-ta Hsu.
\newblock {Extending Shadow Matching to Tightly-Coupled GNSS/INS Integration
  System}.
\newblock \emph{IEEE Transactions on Vehicular Technology}, 69\penalty0
  (5):\penalty0 4979--4991, may 2020.
\newblock ISSN 0018-9545.
\newblock \doi{10.1109/TVT.2020.2981093}.
\newblock URL \url{https://ieeexplore.ieee.org/document/9040648/}.

\bibitem[Liu et~al.(2014)Liu, Liang, Morton, Closas, Zhang, and Hong]{Liu2014a}
Xiaoli Liu, Muqing Liang, Yu~Morton, Pau Closas, Ting Zhang, and Zhigang Hong.
\newblock {Performance evaluation of MSK and OFDM modulations for future GNSS
  signals}.
\newblock \emph{GPS Solutions}, 18\penalty0 (2):\penalty0 163--175, 2014.
\newblock ISSN 15211886.
\newblock \doi{10.1007/s10291-014-0368-6}.

\bibitem[Cimini(1985)]{Cimini1985}
L.~Cimini.
\newblock {Analysis and Simulation of a Digital Mobile Channel Using Orthogonal
  Frequency Division Multiplexing}.
\newblock \emph{IEEE Transactions on Communications}, 33\penalty0 (7):\penalty0
  665--675, jul 1985.
\newblock ISSN 0096-2244.
\newblock \doi{10.1109/TCOM.1985.1096357}.
\newblock URL \url{http://ieeexplore.ieee.org/document/1096357/}.

\bibitem[Koelemeij et~al.(2022)Koelemeij, Dun, Diouf, Dierikx, Janssen, and
  Tiberius]{Koelemeij2022}
Jeroen C.~J. Koelemeij, Han Dun, Cherif E.~V. Diouf, Erik~F. Dierikx, Gerard
  J.~M. Janssen, and Christian C. J.~M. Tiberius.
\newblock {A hybrid optical–wireless network for decimetre-level terrestrial
  positioning}.
\newblock \emph{Nature}, 611\penalty0 (7936):\penalty0 473--478, nov 2022.
\newblock ISSN 0028-0836.
\newblock \doi{10.1038/s41586-022-05315-7}.
\newblock URL \url{https://www.nature.com/articles/s41586-022-05315-7}.

\bibitem[Krasner and McBurney(2022)]{Krasner2022}
Norman Krasner and Paul McBurney.
\newblock {Application of Super Resolution Correlation to Multipath Mitigation
  in an L5 Channel}.
\newblock In \emph{Proceedings of the 35th International Technical Meeting of
  the Satellite Division of The Institute of Navigation (ION GNSS+ 2022)},
  pages 3249--3269, oct 2022.
\newblock \doi{10.33012/2022.18584}.
\newblock URL
  \url{https://www.ion.org/publications/abstract.cfm?articleID=18584}.

\bibitem[Luo et~al.(2021{\natexlab{a}})Luo, Hsu, and Pan]{Luo2021sra}
Yiran Luo, Li-Ta Hsu, and Yi~Pan.
\newblock {A Super-Resolution Algorithm with FRFT Towards GNSS TOA Estimation
  for Multipath Channel}.
\newblock In \emph{Proceedings of the ION GNSS+ 2021, St. Louis, Missouri, USA,
  Sep 20-24}, pages 3350--3359, oct 2021{\natexlab{a}}.
\newblock \doi{10.33012/2021.17951}.
\newblock URL
  \url{https://www.ion.org/publications/abstract.cfm?articleID=17951}.

\bibitem[{Da Rosa Zanatta} et~al.(2020){Da Rosa Zanatta}, {Da Costa}, Antreich,
  Haardt, Elger, {Lopes De Mendonca}, and {De Sousa}]{DaRosaZanatta2020}
Mateus {Da Rosa Zanatta}, Joao Paulo Carvalho~Lustosa {Da Costa}, Felix
  Antreich, Martin Haardt, Gordon Elger, Fabio~Lucio {Lopes De Mendonca}, and
  Rafael~Timoteo {De Sousa}.
\newblock {Tensor-Based Framework With Model Order Selection and High Accuracy
  Factor Decomposition for Time-Delay Estimation in Dynamic Multipath
  Scenarios}.
\newblock \emph{IEEE Access}, 8:\penalty0 174931--174942, 2020.
\newblock ISSN 2169-3536.
\newblock \doi{10.1109/ACCESS.2020.3024597}.
\newblock URL \url{https://ieeexplore.ieee.org/document/9200326/}.

\bibitem[Luo and El-Sheimy(2022)]{Luo2022}
Yiran Luo and Naser El-Sheimy.
\newblock {Improving GNSS Positioning by De-noising Consecutive Correlator
  Outputs Using Graph Fourier Transform Filtering}.
\newblock In \emph{Proceedings of the ION GNSS+ 2022, Denver, Colorado, USA,
  September 19-23}, pages 3340--3348, sep 2022.
\newblock \doi{10.33012/2022.18397}.
\newblock URL
  \url{https://www.ion.org/publications/abstract.cfm?articleID=18397}.

\bibitem[Suzuki et~al.(2020)Suzuki, Matsuo, and Amano]{Suzuki2020}
Taro Suzuki, Kazuki Matsuo, and Yoshiharu Amano.
\newblock {Rotating GNSS Antennas: Simultaneous LOS and NLOS Multipath
  Mitigation}.
\newblock \emph{GPS Solutions}, 24\penalty0 (3):\penalty0 86, jul 2020.
\newblock ISSN 15211886.
\newblock \doi{10.1007/s10291-020-01006-w}.
\newblock URL \url{https://doi.org/10.1007/s10291-020-01006-w
  http://link.springer.com/10.1007/s10291-020-01006-w}.

\bibitem[Hong et~al.(2020)Hong, Wang, Chang, and Yin]{Hong2020}
Xi~Hong, Wenjie Wang, Ning Chang, and Qinye Yin.
\newblock {A subspace-based code tracking loop design for GPS multi-antenna
  receiver in multipath environment}.
\newblock \emph{GPS Solutions}, 24\penalty0 (4):\penalty0 109, oct 2020.
\newblock ISSN 1080-5370.
\newblock \doi{10.1007/s10291-020-01020-y}.
\newblock URL \url{https://link.springer.com/10.1007/s10291-020-01020-y}.

\bibitem[Daneshmand et~al.(2013)Daneshmand, Broumandan, Sokhandan, and
  Lachapelle]{Daneshmand2013}
S.~Daneshmand, A.~Broumandan, N.~Sokhandan, and G.~Lachapelle.
\newblock {GNSS multipath mitigation with a moving antenna array}.
\newblock \emph{IEEE Transactions on Aerospace and Electronic Systems},
  49\penalty0 (1):\penalty0 693--698, 2013.
\newblock ISSN 00189251.
\newblock \doi{10.1109/TAES.2013.6404136}.

\bibitem[Lau and Cross(2007)]{Lau2007}
Lawrence Lau and Paul Cross.
\newblock {Development and testing of a new ray-tracing approach to GNSS
  carrier-phase multipath modelling}.
\newblock \emph{Journal of Geodesy}, 81\penalty0 (11):\penalty0 713--732, oct
  2007.
\newblock ISSN 0949-7714.
\newblock \doi{10.1007/s00190-007-0139-z}.
\newblock URL \url{http://link.springer.com/10.1007/s00190-007-0139-z}.

\bibitem[Yan et~al.(2022)Yan, Chen, Tang, and Zhu]{Yan2022}
Zhe Yan, Xiyuan Chen, Xinhua Tang, and Xuefen Zhu.
\newblock {Design and Performance Evaluation of the Improved INS-Assisted
  Vector Tracking for the Multipath in Urban Canyons}.
\newblock \emph{IEEE Transactions on Instrumentation and Measurement}, 71,
  2022.
\newblock ISSN 15579662.
\newblock \doi{10.1109/TIM.2022.3204107}.

\bibitem[Smolyakov et~al.(2020)Smolyakov, Rezaee, and Langley]{Smolyakov2020}
Ivan Smolyakov, Mohammad Rezaee, and Richard~B. Langley.
\newblock {Resilient multipath prediction and detection architecture for
  low-cost navigation in challenging urban areas}.
\newblock \emph{Navigation, Journal of the Institute of Navigation},
  67\penalty0 (2):\penalty0 397--409, jun 2020.
\newblock ISSN 00281522.
\newblock \doi{10.1002/navi.362}.

\bibitem[{Spilker Jr}(1996)]{Parkinson1996}
James~J {Spilker Jr}.
\newblock {Fundamentals of signal tracking theory}.
\newblock In Bradford~W. Parkinson, James~J {Spilker Jr}, Penina Axelrad, and
  Per Enge, editors, \emph{Global Positioning System: Theory And Applications,
  Volume 1}, chapter~7. American Institute of Aeronautics and Astronautics,
  Inc., Washington, DC, 1996.

\bibitem[Zhodzishsky et~al.(1998)Zhodzishsky, Yudanov, Veitsel, and
  Ashjaee]{Zhodzishsky1998}
M~Zhodzishsky, S~Yudanov, V~Veitsel, and J~Ashjaee.
\newblock {Co-OP Tracking for Carrier Phase}.
\newblock In \emph{Proceedings of the ION GPS 1998, Nashville, USA, Sep 15-18},
  pages 653--664, 1998.

\bibitem[Henkel et~al.(2009)Henkel, Giger, and Gunther]{Henkel2009}
Patrick Henkel, Kaspar Giger, and Christoph Gunther.
\newblock {Multifrequency, Multisatellite Vector Phase-Locked Loop for Robust
  Carrier Tracking}.
\newblock \emph{IEEE Journal of Selected Topics in Signal Processing},
  3\penalty0 (4):\penalty0 674--681, aug 2009.
\newblock ISSN 1932-4553.
\newblock \doi{10.1109/JSTSP.2009.2025637}.
\newblock URL \url{http://ieeexplore.ieee.org/document/5166551/}.

\bibitem[Shafaati et~al.(2018)Shafaati, Lin, Broumandan, and
  Lachapelle]{Shafaati2018a}
Ahmad Shafaati, Tao Lin, Ali Broumandan, and G{\'{e}}rard Lachapelle.
\newblock {Design and Implementation of an RTK-Based Vector Phase Locked Loop}.
\newblock \emph{Sensors}, 18\penalty0 (3):\penalty0 845, mar 2018.
\newblock ISSN 1424-8220.
\newblock \doi{10.3390/s18030845}.
\newblock URL \url{http://www.mdpi.com/1424-8220/18/3/845}.

\bibitem[Satyanarayana et~al.(2012)Satyanarayana, Borio, and
  Lachapelle]{Satyanarayana2012}
Shashank Satyanarayana, Daniele Borio, and Gerard Lachapelle.
\newblock {A Composite Model for Indoor GNSS Signals: Characterization,
  Experimental Validation and Simulation}.
\newblock \emph{Navigation}, 59\penalty0 (2):\penalty0 77--92, jun 2012.
\newblock ISSN 00281522.
\newblock \doi{10.1002/navi.8}.
\newblock URL \url{https://onlinelibrary.wiley.com/doi/10.1002/navi.8}.

\bibitem[Kelly and Braasch(2001)]{Kelly2001}
J.M. Kelly and M.S. Braasch.
\newblock {Validation of theoretical GPS multipath bias characteristics}.
\newblock In \emph{2001 IEEE Aerospace Conference Proceedings (Cat.
  No.01TH8542)}, volume~3, pages 3/1317--3/1325. IEEE, 2001.
\newblock ISBN 0-7803-6599-2.
\newblock \doi{10.1109/AERO.2001.931362}.
\newblock URL \url{http://ieeexplore.ieee.org/document/931362/}.

\bibitem[{VAN NEE}(1992)]{VANNEE1992}
RICHARD D.~J. {VAN NEE}.
\newblock {Multipath Effects on GPS Code Phase Measurements}.
\newblock \emph{Navigation}, 39\penalty0 (2):\penalty0 177--190, jun 1992.
\newblock ISSN 00281522.
\newblock \doi{10.1002/j.2161-4296.1992.tb01873.x}.
\newblock URL
  \url{https://onlinelibrary.wiley.com/doi/10.1002/j.2161-4296.1992.tb01873.x}.

\bibitem[Kaplan and Hegarty(2017)]{Kaplan2017}
Elliott~D. Kaplan and Christopher Hegarty.
\newblock \emph{{Understanding GPS/GNSS. Principles and applications}}.
\newblock Artech house, 3rd edition, 2017.
\newblock ISBN 1580538940.
\newblock \doi{10.1016/s1364-6826(97)83337-8}.

\bibitem[Lashley(2009)]{Lashley2009a}
Matthew Lashley.
\newblock \emph{{Modelling and performance analysis of GPS vector tracking
  algorithms}}.
\newblock Ph.d. thesis, Auburn University, Auburn, USA, 2009.
\newblock URL
  \url{http://ezproxy.lib.ucalgary.ca/login?url=https://search.proquest.com/docview/304830161?accountid=9838}.

\bibitem[DIetmayer et~al.(2020)DIetmayer, Kunzi, Garzia, Overbeck, and
  Felber]{DIetmayer2020}
Katrin DIetmayer, Florian Kunzi, Fabio Garzia, Matthias Overbeck, and Wolfgang
  Felber.
\newblock {Real Time Results of Vector Delay Lock Loop in a Light Urban
  Scenario}.
\newblock \emph{2020 IEEE/ION Position, Location and Navigation Symposium,
  PLANS 2020}, pages 1230--1236, 2020.
\newblock \doi{10.1109/PLANS46316.2020.9109832}.

\bibitem[Luo et~al.(2022{\natexlab{a}})Luo, Hsu, Zhang, and
  El-Sheimy]{Luo2022aa}
Yiran Luo, Li-Ta Hsu, Zhetao Zhang, and Naser El-Sheimy.
\newblock {Improving GNSS baseband using an RTK-position-aided code tracking
  algorithm}.
\newblock \emph{GPS Solutions}, 26\penalty0 (4):\penalty0 125, oct
  2022{\natexlab{a}}.
\newblock ISSN 1080-5370.
\newblock \doi{10.1007/s10291-022-01305-4}.
\newblock URL \url{https://link.springer.com/10.1007/s10291-022-01305-4}.

\bibitem[Groves(2013)]{Groves2013}
Paul~D Groves.
\newblock \emph{{Principles of GNSS, inertial, and multisensor integrated
  navigation systems}}.
\newblock Artech House, Boston{\&}London, 2nd edition, 2013.
\newblock ISBN 1608070050.

\bibitem[Harke and O'Keefe(2022)]{Harke2022}
Kelly Harke and Kyle O'Keefe.
\newblock {Gyroscope Drift Estimation of a GPS/MEMSINS Smartphone Sensor
  Integration Navigation System for Kayaking}.
\newblock In \emph{Proceedings of the ION GNSS+ 2022}, pages 1413--1427,
  Denver, the US, oct 2022.
\newblock \doi{10.33012/2022.18324}.

\bibitem[Zhang et~al.(2022)Zhang, Liu, Chen, Feng, and Niu]{Zhang2022}
Tisheng Zhang, Shan Liu, Qijin Chen, Xin Feng, and Xiaoji Niu.
\newblock {Carrier-Phase-Based Initial Heading Alignment for Land Vehicular
  MEMS GNSS/INS Navigation System}.
\newblock \emph{IEEE Transactions on Instrumentation and Measurement},
  71:\penalty0 1--13, 2022.
\newblock ISSN 0018-9456.
\newblock \doi{10.1109/TIM.2022.3208646}.
\newblock URL \url{https://ieeexplore.ieee.org/document/9899471/}.

\bibitem[Luo et~al.(2021{\natexlab{b}})Luo, Hsu, Xiang, Xu, and Yu]{Luo2021apa}
Yiran Luo, Li-Ta Hsu, Yan Xiang, Bing Xu, and Chunyang Yu.
\newblock {An Absolute-Position-Aided Code Discriminator Towards GNSS Receivers
  for Multipath Mitigation}.
\newblock In \emph{Proceedings of the ION GNSS+ 2021, St. Louis, Missouri, USA,
  Sep 20-24}, pages 3772--3782, 2021{\natexlab{b}}.
\newblock \doi{10.33012/2021.18001}.
\newblock URL
  \url{https://www.ion.org/publications/abstract.cfm?articleID=18001}.

\bibitem[Lashley et~al.(2021)Lashley, Martin, and Sennott]{Lashley2021}
Matthew~V. Lashley, Scott Martin, and James Sennott.
\newblock {Vector Processing}.
\newblock In Y.~T.~Jade Morton, Frank van Diggelen, James~J. {Spilker Jr.},
  Bradford~W. Parkinson, and Grace Lo, ShermanGao, editors, \emph{Position,
  Navigation, and Timing Technologies in the 21st Century: Integrated Satellite
  Navigation, Sensor Systems, and Civil Applications}, volume~1. John Wiley
  {\&} Sons, 2021.

\bibitem[Lashley and Bevly(2013)]{Lashley2013a}
Matthew Lashley and David~M. Bevly.
\newblock {Performance comparison of deep integration and tight coupling}.
\newblock \emph{Navigation, Journal of the Institute of Navigation},
  60\penalty0 (3):\penalty0 159--178, sep 2013.
\newblock ISSN 00281522.
\newblock \doi{10.1002/navi.43}.
\newblock URL \url{https://onlinelibrary.wiley.com/doi/10.1002/navi.43}.

\bibitem[Herrera et~al.(2016)Herrera, Suhandri, Realini, Reguzzoni, and
  de~Lacy]{Herrera2016}
Antonio~M. Herrera, Hendy~F. Suhandri, Eugenio Realini, Mirko Reguzzoni, and
  M.~Clara de~Lacy.
\newblock {goGPS: open-source MATLAB software}.
\newblock \emph{GPS Solutions}, 20\penalty0 (3):\penalty0 595--603, 2016.
\newblock ISSN 15211886.
\newblock \doi{10.1007/s10291-015-0469-x}.

\bibitem[Luo et~al.(2019{\natexlab{a}})Luo, Yu, Xu, Li, Tsai, Li, and
  El-Sheimy]{Luo2019}
Yiran Luo, Chunyang Yu, Bing Xu, Jian Li, Guang-Je Tsai, You Li, and Naser
  El-Sheimy.
\newblock {Assessment of Ultra-Tightly Coupled GNSS/INS Integration System
  towards Autonomous Ground Vehicle Navigation Using Smartphone IMU}.
\newblock In \emph{2019 IEEE International Conference on Signal, Information
  and Data Processing (ICSIDP)}, pages 1--6. IEEE, dec 2019{\natexlab{a}}.
\newblock ISBN 978-1-7281-2345-5.
\newblock \doi{10.1109/ICSIDP47821.2019.9173292}.
\newblock URL \url{https://ieeexplore.ieee.org/document/9173292/}.

\bibitem[Luo et~al.(2021{\natexlab{c}})Luo, Li, Wang, and
  El-Sheimy]{Luo2021utc}
Yiran Luo, You Li, Jin Wang, and Naser El-Sheimy.
\newblock {Supporting GNSS Baseband Using Smartphone IMU and Ultra-Tight
  Integration}.
\newblock \emph{arXiv preprint arXiv:2111.02613}, 2021{\natexlab{c}}.
\newblock URL \url{http://arxiv.org/abs/2111.02613}.

\bibitem[Luo et~al.(2019{\natexlab{b}})Luo, Li, Yu, Lyu, Yue, and
  El-Sheimy]{Luo2019b}
Yiran Luo, Jian Li, Chunyang Yu, Zhitao Lyu, Zhe Yue, and Naser El-Sheimy.
\newblock {A GNSS software-defined receiver with vector tracking techniques for
  land vehicle navigation}.
\newblock In \emph{Proc. ION 2019 Pacific PNT Meeting, Honolulu, Hawaii, USA,
  April 8-11}, volume 2019-April, pages 713--727, 2019{\natexlab{b}}.
\newblock ISBN 0936406224.
\newblock \doi{10.33012/2019.16834}.
\newblock URL
  \url{https://www.ion.org/publications/abstract.cfm?articleID=16834}.

\bibitem[Faragher(2012)]{Faragher2012}
Ramsey Faragher.
\newblock {Understanding the Basis of the Kalman Filter Via a Simple and
  Intuitive Derivation [Lecture Notes]}.
\newblock \emph{IEEE Signal Processing Magazine}, 29\penalty0 (5):\penalty0
  128--132, sep 2012.
\newblock ISSN 1053-5888.
\newblock \doi{10.1109/MSP.2012.2203621}.
\newblock URL \url{http://ieeexplore.ieee.org/document/6279585/}.

\bibitem[Takasu and Yasuda(2009)]{Takasu2009}
Tomoji Takasu and Akio Yasuda.
\newblock {Development of the low-cost RTK-GPS receiver with an open source
  program package RTKLIB}.
\newblock In \emph{Proceedings of the International symposium on GPS/GNSS,
  Jeju, Korea}, pages 4--6, 2009.

\bibitem[Luo et~al.(2022{\natexlab{b}})Luo, Hsu, and El-Sheimy]{Luo2022up}
Yiran Luo, Li-Ta Hsu, and Naser El-Sheimy.
\newblock {A Baseband MLE for Snapshot GNSS Receiver Using Super-Long-Coherent
  Correlation in a Fractional Fourier Domain}.
\newblock \emph{unpublished}, 2022{\natexlab{b}}.

\bibitem[Fern{\'{a}}ndez-Hern{\'{a}}ndez
  et~al.(2022)Fern{\'{a}}ndez-Hern{\'{a}}ndez, L{\'{o}}pez-Salcedo, and
  Seco-Granados]{Fernandez-Hernandez2022}
Ignacio Fern{\'{a}}ndez-Hern{\'{a}}ndez, Jos{\'{e}}~A L{\'{o}}pez-Salcedo, and
  Gonzalo Seco-Granados.
\newblock {9 Snapshot Receivers}.
\newblock In Kai Borre, Ignacio Fern{\'{a}}ndez-Hern{\'{a}}ndez, Jos{\'{e}}~A.
  L{\'{o}}pez-Salcedo, and M.~Zahidul~H. Bhuiyan, editors, \emph{GNSS Software
  Receivers}. Cambridge University Press, oct 2022.
\newblock ISBN 9781108934176.
\newblock \doi{10.1017/9781108934176}.

\bibitem[Tsang et~al.(2022)Tsang, Luo, and Hsu]{Tsang2022}
Chin~Lok Tsang, Yiran Luo, and Li-Ta Hsu.
\newblock {Long Coherent Open-Loop GPS L5Q Signal Positioning: A Case Study for
  an Urban Area in Hong Kong}.
\newblock In \emph{Proceedings of the ION GNSS+ 2022, Denver, Colorado, USA,
  September 19-23}, pages 2025--2041, oct 2022.
\newblock \doi{10.33012/2022.18326}.

\bibitem[van Graas et~al.(2009)van Graas, Soloviev, {Uijt de Haag}, and
  Gunawardena]{VanGraas2009}
Frank van Graas, Andrey Soloviev, Maarten {Uijt de Haag}, and Sanjeev
  Gunawardena.
\newblock {Closed-loop sequential signal processing and open-loop batch
  processing approaches for GNSS receiver design}.
\newblock \emph{IEEE Journal on Selected Topics in Signal Processing},
  3\penalty0 (4):\penalty0 571--586, 2009.
\newblock ISSN 19324553.
\newblock \doi{10.1109/JSTSP.2009.2023350}.

\end{thebibliography}






\end{document}